\documentclass[a4paper,11pt]{article}
\pdfoutput=1 % if your are submitting a pdflatex (i.e. if you have
\usepackage{jheppub} % for details on the use of the package, please
\usepackage{rotating}
\usepackage{multicol}
\usepackage{graphicx}
\usepackage{booktabs}
\usepackage{amssymb,bm,mathrsfs,bbm,amscd}
\usepackage{subfig,color,slashed}
\usepackage{amssymb}
\usepackage[normalem]{ulem}
\usepackage[utf8]{inputenc}
\usepackage{ulem}
\usepackage{kotex}
\usepackage{lscape}
\usepackage{lipsum}

\def\lsim{\mathrel{\rlap{\lower4pt\hbox{\hskip1pt$\sim$}}
    \raise1pt\hbox{$<$}}}         %less than or approx. symbol
\def\gsim{\mathrel{\rlap{\lower4pt\hbox{\hskip1pt$\sim$}}
    \raise1pt\hbox{$>$}}}         %greater than or approx. symbol

\title{Search for muon-philic new light gauge boson \\ at Belle II}

\author[\dagger]{Yongsoo Jho,}
\author[\dagger]{Youngjoon Kwon,}
\author[\dagger]{Seong Chan Park}
\author[\star]{and Po-Yan Tseng}

\affiliation[\dagger]{Department of Physics and IPAP, Yonsei University, \\
Seoul 03722, Republic of Korea}
\affiliation[\star]{Kavli IPMU (WPI), UTIAS, The University of Tokyo, \\
Kashiwa, Chiba 277-8583, Japan}

\emailAdd{jys34@yonsei.ac.kr}
\emailAdd{yjkwon63@yonsei.ac.kr}
\emailAdd{sc.park@yonsei.ac.kr}
\emailAdd{tpoyan1209@gmail.com}

\abstract{
Motivated by the long-lasting $3.5\sigma$ discrepancy in the anomalous magnetic moment of muon, we consider a new muon-specific force mediated by a light gauge boson, $X$, with mass $m_X < 2m_\mu$ and the coupling constant $g_X \sim (10^{-4}, 10^{-3})$. We show that the Belle~II experiment has a robust chance to probe such a light boson in $e^+ e^- \to \mu^+ \mu^- + X$ channel and cover the most interesting parameter space explaining the discrepancy with the planned target luminosity, $\int dt \ {\cal L}=50~{\rm ab^{-1}}$. The clean signal of muon-pair plus missing energy at Belle~II can be a smoking gun for the new gauge boson. We expect that the (invisibly decaying) muon-philic light ($m_X \lsim 2 m_\mu$) gauge boson can be probed down to $g_X \gsim 1.5 \times 10^{-4} \ (4.6 \times 10^{-4}, \ 2.3 \times 10^{-4})$ for 50 (1, 10) ab${}^{-1}$ search.
}
\begin{document} 
\maketitle
\flushbottom

%%%%%%%%%%%%%%%%%%%%%%%%
\section{Introduction}
%%%%%%%%%%%%%%%%%%%%%%%%

After the Higgs discovery in 2012, we are now entering the new era of particle physics. The main goal now is to uncover physics beyond the standard model (SM) even though there are still more rooms to improve the precision of the measurements especially in the Higgs quartic and cubic couplings as well as the top quark (pole) mass, which are crucial to determine the stability of our universe \cite{Degrassi:2012ry, Buttazzo:2013uya}.\footnote{Also see \cite{Hamada:2014iga, Hamada:2014wna} in the context of cosmological Higgs inflation.}

Even without any theoretical prejudice, we are actually facing the observational problems, which enforce us to modify or enlarge the standard model. In particular, the significant discrepancy in the anomalous magnetic moment of the muon remains one of the largest anomalies in particle physics~\cite{Blum:2018mom, Tanabashi:2018oca, Brown:2001mga, Bennett:2004pv}:
\begin{eqnarray}
a_\mu^{\rm exp} -a_\mu^{\rm SM} & = & (268\pm 63_{\rm exp} \pm 43_{\rm the}) \times 10^{-11},
\label{eq:ammm}
\end{eqnarray}
where the errors are from experiment and theory prediction, respectively. Many well motivated theoretical solutions to fit the data have been proposed~\cite{Moroi:1995yh, Pospelov:2008zw, Czarnecki:2001pv, Park:2003sq, Park:2001uc} but no one has been experimentally confirmed so far~\cite{Jegerlehner:2009ry}. 

It is well-known that light, weakly coupled particles can bring theoretical predictions into agreement with observations \cite{Pospelov:2008zw}. With a simplified interaction with muon, ${\cal L}=-g_X X_\mu \bar{\mu}\gamma^\mu \mu$,  the light ($m_X \lsim 2m_\mu$) gauge boson ($X_\mu$)  contribution to the anomalous magnetic moment of muon at one-loop level is
\begin{eqnarray}
\Delta a_\mu^X & = & \frac{g_X^2}{8\pi^2} \int_0^1 dz \  \frac{2z(1-z)^2}{(1-z)^2 + (m_X/m_\mu)^2 z} \ .
\end{eqnarray}

\begin{figure}[h]
\centering
\includegraphics[width=0.65\textwidth]{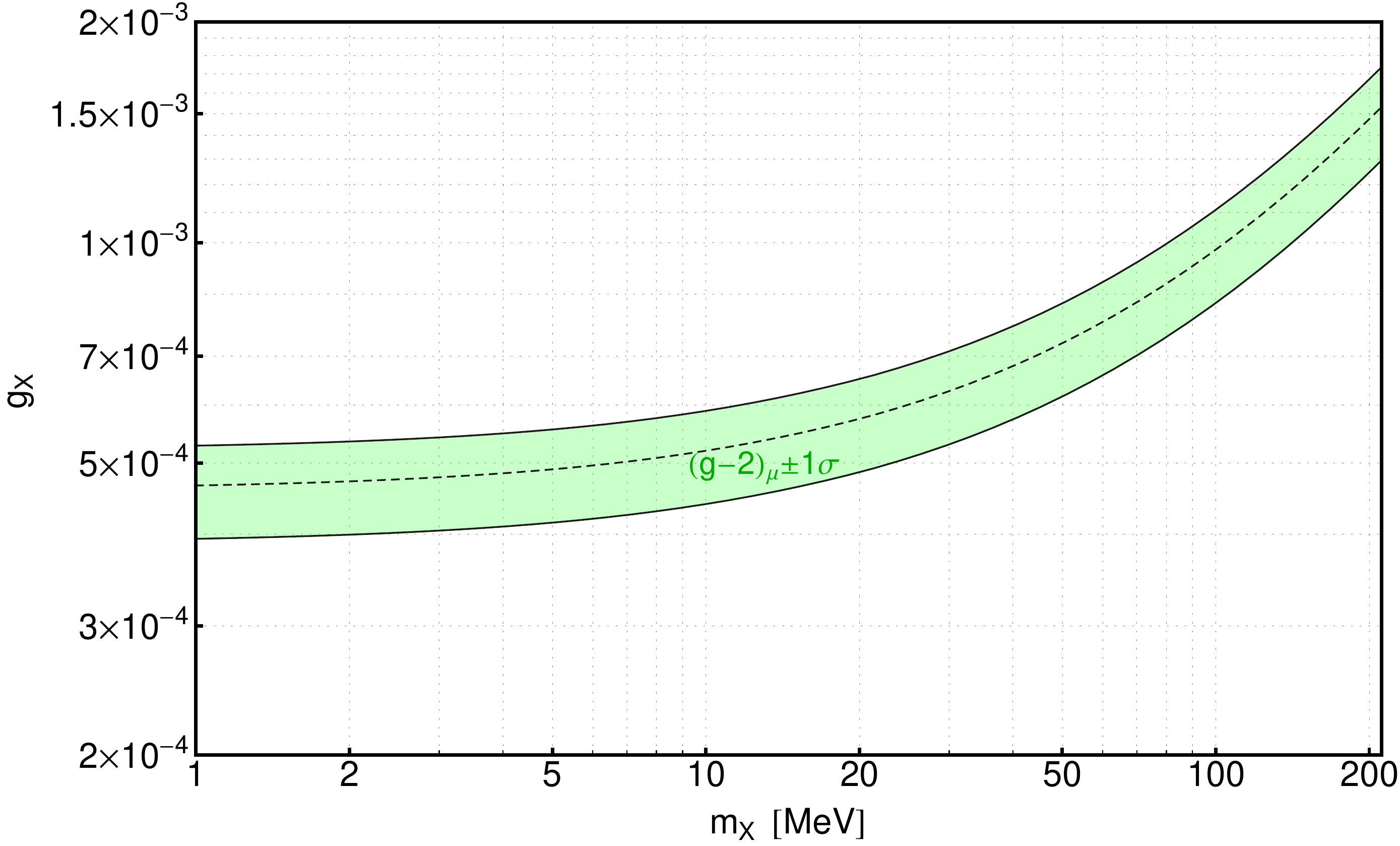}
\caption{The ball park parameter space fitting $\Delta a_\mu$ allowing the $1\sigma$ uncertainty. 
The $1\sigma$ uncertainty is estimated by $\Delta \sigma = \sqrt{63^2+43^2} \times 10^{-11}$.}
\label{fig:ammm}
\end{figure}

The integration is easily done numerically and found to be positive and close to unity when $m_X/m_\mu \lsim 1$ so that $\Delta a_\mu^X \sim g_X^2/8\pi^2 \sim 3\times 10^{-9}$. Hence $g_X\sim 5 \times 10^{-4}$ is desired. This sets up the ball-park range of parameters for our study. (see Fig.~\ref{fig:ammm})~\footnote{A light ($<$ GeV) dark photon with kinetic mixing $\epsilon \sim 10^{-3}$  and flavor-universal couplings, has been ruled out for either cases where it decays to visible final-states only, or to invisible final states only \cite{Battaglieri:2017aum}. However, the partially visible and partially invisible decays of dark photon scenario is currently still allowed. Therefore, future sensitivities from Belle II monophoton search and BABAR displaced track re-analysis will probe this region \cite{Mohlabeng:2019vrz}.} 
%%%%%%%%%%%%%%%%%%%%%%%%
%%%%%%%%%%%%%%%%%%%%%%%%

When we target to the new light gauge boson, $X_\mu$, we don't really need a huge center-of-mass frame energy of the LHC or other future experiments but rather a precise measurement at a relatively low energy experiment. In this letter, we would focus on the Belle~II experiment~\cite{Abe:2010gxa}, which has been just started and will get scientific data in coming years~\cite{Shwartz:2015kja}. Indeed, as we will show in detail, the Belle~II experiment would be an ideal place for our purpose.

Most dark photon searches at low-energy colliders have considered the mono-photon process $e^- e^+ \rightarrow \gamma A'$ which depends on the kinetic mixing $\epsilon_{\gamma A'}$ between the Standard Model photon and the dark photon $A'$ \cite{Lees:2014xha,Lees:2017lec}. For the muonic force such as gauged $L_\mu - L_\tau$ \cite{He:1991qd}, the similar mono-photon channel has been considered for `minimal' gauged $L_\mu - L_\tau$ whose kinetic mixing is induced by only SM $\mu$ and $\tau$ loops \cite{Kaneta:2016uyt,Araki:2017wyg}. To be specific for muonic force, we have considered the $X$-bremsstrahlung process $e^- e^+ \rightarrow \mu^- \mu^+ X$, $X \rightarrow $ (invisible) in the muon pair production, which is independent on the kinetic mixing $\epsilon_{\gamma X}$.

This paper is organized as follows. In the next section (Sec.~\ref{sec:model}) we first set up our theoretical model, a minimal model of muon-philic gauge boson, where the necessary interactions and the most relevant parameters are introduced. We are taking the anomaly-free condition into account for consistency while requiring the model to remain minimal. In Sec.~\ref{sec:BELLE}, we study the signature at Belle~II experiment in $e^+e^-\to \mu^+\mu^- X$ channel then optimize the signal/background taking the spectral shape and the missing transverse energy $\slashed{E}_T$ and the missing mass $m_{\rm miss}^2$ cuts of muons into account. We show the potential coverage of the Belle~II experiment in comparison with other relevant experiments. We finally conclude in Sec.~\ref{sec:conclusion}.

%%%%%%%%%%%%%%%%%%%%%%%%
\section{Model \label{sec:model}}
%%%%%%%%%%%%%%%%%%%%%%%%

To incorporate the muonic new force for muon-philic new gauge boson, we extend the SM by including a new $U(1)_X$ gauge symmetry. The Lagrangian now contains the kinetic term, mass term and the gauge interaction term for the gauge boson, $X_\mu$, of the new gauge symmetry:
\begin{eqnarray}
\mathcal{L} & \supset & \mathcal{L}_{\rm SM} - \frac{1}{4} X_{\mu\nu} X^{\mu\nu} - \frac{1}{2} m_X^2 X_\mu X^\mu - g_X X_\mu J^\mu_X,
\end{eqnarray}
where $X_{\mu\nu} =\partial_\mu X_\nu -\partial_\nu X_\mu$ denotes the field strength tensor of the new gauge interaction and $g_X$ is the gauge coupling constant. The $U(1)_X$ current is given by the charge assignment of the SM fields (and extra fields too, in principle). The kinetic mixing between $U(1)_X$ and SM $U(1)_Y$ gauge bosons induce the small electromagnetic current contribution $\sim \epsilon_{\gamma X} J_{\rm EM}^\mu$ but we do not focus on it since they are much more suppressed by both $\epsilon_{\gamma X}$ and $g_X$.

As a simple but consistent example, we may take the leptonic symmetry, $X=(L_\mu-L_\tau)$, which is  anomaly free. In this case, the new gauge boson couples with the muonic  and tauonic  currents with their corresponding (left-chiral) neutrinos~\cite{He:1991qd, Kahn:2018cqs}:
\begin{eqnarray}
J_X^\mu & = & \bar{\mu} \gamma^\mu \mu - \bar{\tau} \gamma^\mu \tau + \bar{\nu}_\mu \gamma^\mu  \nu_{\mu L} - \bar{\nu}_\tau \gamma^\mu \nu_{\tau L}.
\label{eq:jmu}
\end{eqnarray}

It is important to notice that as long as the new boson is light below the muonic threshold, $m_X \lsim 2 m_\mu \approx 2 \times 105.7$ MeV,  the $X_\mu$ boson would decay mainly to neutrinos (i.e. $\nu_\mu\bar{\nu}_\mu$, $\nu_\tau\bar{\nu}_\tau$) because all other channels are kinematically forbidden. 

It may be worth considering other potentially interesting options free from anomaly.  The first, seemingly minimal, option is the solely muonic symmetry $U(1)_{L_\mu}$, which couples to only muon and muon-neutrino at low-energies. This option looks indeed good enough for phenomenological studies of muonic force.  However, as pointed out in \cite{Dror:2017ehi, Dror:2017nsg}, regardless of the UV structure (content of anomaly-cancelling fermion), they would be strongly constrained by Wess-Zumino counterterm contributions to exotic $Z \rightarrow \gamma X$ decays \cite{Acciarri:1997im} from the 4-dimension operator $g_X g'^2 \epsilon^{\mu \nu \rho \sigma} X_\mu B_\nu \partial_\rho B_\sigma$ and FCNC processes such as $B \rightarrow KX$, $K \rightarrow \pi X$ \cite{Grygier:2017tzo, Artamonov:2008qb} from the other operator $g_X g^2 \epsilon^{\mu \nu \rho \sigma} X_\mu (W_\nu^a \partial_\rho W_\sigma^a + \frac{1}{3} g \epsilon^{abc} W_\nu^a W_\rho^b W_\sigma^c)$. Another potentially interesting option for UV completion free from anomaly is $U(1)_{L_\mu - B_{i=1,2,3}}$, which would open not only leptonic but also hadronic interactions. This case is also highly constrained by e.g. proton beam-dump experiment \cite{Bergsma:1985qz}.\footnote{The proton beam-dump experiment usually use the proton bremsstrahlung $p N \rightarrow p N X$ \cite{Blumlein:2013cua} and the meson decay process, such as $\pi^0 \rightarrow X \gamma$ \cite{Blumlein:2011mv} and $\eta \rightarrow X \gamma$ \cite{Gninenko:2012eq}, to constrain $B_1$ (the baryon number for the first generation) mainly. $B_2$ and $B_3$ (i.e. the baryon number for second and third generation, respectively.) still can be free from this kind of low energy constraints unless we consider the large kinetic mixing with $U(1)_Y$ gauge boson.}
Thus, to avoid unnecessary complication in our analysis, we will focus on the $U(1)_{L_\mu-L_\tau}$ case below. 

In addition, one can naturally extend the list of interactions mediated by muon-philic $X$ gauge boson, including dark sector particles. It provides possible scenarios of light dark matter at sub-GeV scale \cite{Foldenauer:2018zrz}. If one considers additional interactions between $X$ and the dark sector particles, $N_\chi$ (vector-like) fermions $\chi_i$ for example, as
\begin{eqnarray}
\mathcal{L} & = & \mathcal{L}_{\rm minimal} + \sum_{i=1}^{N_\chi} \Bigl [ \bar{\chi}_i (i \slashed{\partial} - m_{\chi,i}) \chi_i - g_D ( \bar{\chi}_i \gamma^\mu \chi_i ) X_\mu \Bigr ],
\end{eqnarray}
the width of $X$ boson can be enhanced as $\Gamma_{X,\text{total}} = (1+ \delta_{\rm NM}) \cdot \Gamma_{\rm Minimal}$ where
\begin{eqnarray}
\delta_{\rm NM} & = & \sum_{i=1}^{N_\chi} \frac{g_D^2}{g_X^2} \Bigl (1+ \frac{m_{\chi_i}^2}{m_X^2} \Bigr ) \sqrt{1- 4 \frac{m_{\chi_i}^2}{m_X^2}}
\end{eqnarray}
and $\Gamma_{\rm Minimal} = m_X g_X^2 / 12\pi$ is the total width of minimal gauged $L_\mu - L_\tau$ case. $N_\chi$ is the number of fermion species in the dark sector.

\begin{figure}[h]
\centering
\includegraphics[width=0.9\textwidth]{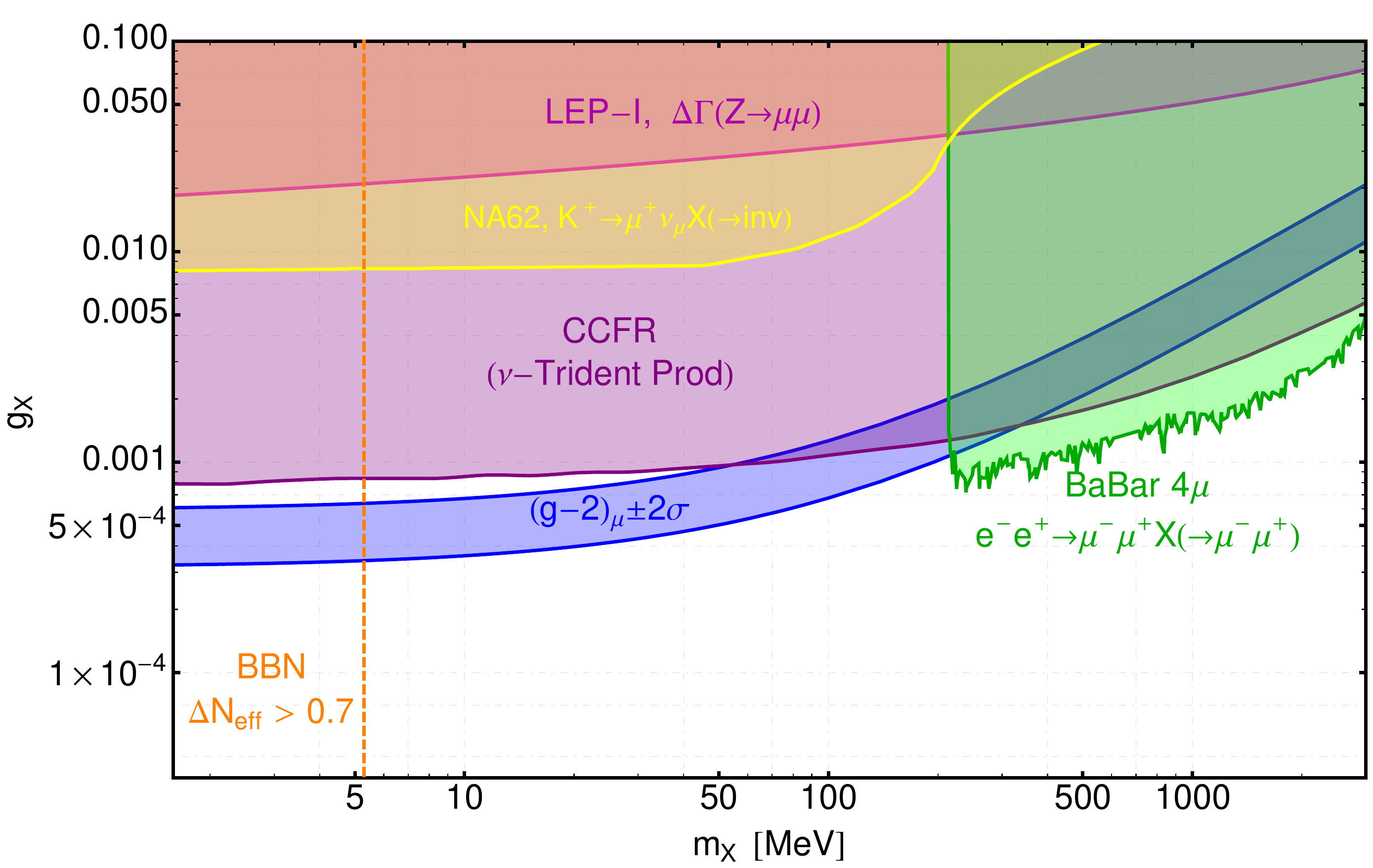}
\caption{The bounds from the previous muon-philic gauge boson searches (purple, magenta, green and yellow), the indirect bound from BBN constraint (orange), and $(g-2)_\mu$ desired parameters (blue). \label{fig_previous_bounds}}
\end{figure}

Before studying the future perspectives of finding the muon-philic new gauge boson at Belle~II experiment, we first consider the existing constraints in the kinematic range of our interest from various experiments as follows:

\begin{itemize}
\item \emph{$Z$-pole precision measurement}. The $X$ boson can contribute to the $Z\mu^+\mu^-$ vertex correction at one-loop level thus modifying the muonic decay width of $Z$ boson by $\Delta \Gamma (Z \rightarrow \mu^- \mu^+)$
\begin{eqnarray}
\frac{\Delta \Gamma(Z \rightarrow \mu^- \mu^+)}{\Gamma(Z \rightarrow \mu^- \mu^+)} & = & \frac{g_X^2}{16\pi^2} F_2 \Bigl ( \frac{m_X^2}{m_Z^2} \Bigr ),
\end{eqnarray}
where the loop-function is 
\begin{eqnarray}
F_2 (x) & \equiv & -2 \left \{ \frac{7}{4} + x + \Bigl ( x + \frac{3}{2} \Bigr ) \ln x + (1+x)^2 \Bigl [ \text{Li}_2 \Bigl ( \frac{x}{1+x} \Bigr ) + \frac{1}{2} \ln^2 \Bigl ( \frac{x}{1+x} \Bigr ) - \frac{\pi^2}{6} \Bigr ] \right \} \nonumber \\
\end{eqnarray}
with the polylogarithmic function of order $2$ being $\text{Li}_2(x)=-\int_0^x \frac{dt}{t}\ln(1-t) $~\cite{Carone:1994aa}. We set the bound for this correction taking the precision measurement at $Z$-pole as
\begin{eqnarray}
\left | \frac{g_X^2}{16\pi^2} F_2 \Bigl ( \frac{m_X^2}{m_Z^2} \Bigr ) \right | & < & \left | \frac{\Gamma (Z \rightarrow \mu^- \mu^+)}{\Gamma (Z \rightarrow e^- e^+)} - 1 \right |
\label{eq:bound1}
\end{eqnarray}
where we used the values \cite{Tanabashi:2018oca}
\begin{eqnarray}
\text{Br} (Z \rightarrow e^- e^+) & = & 3.3632 \pm 0.0042 \%, \\
\text{Br} (Z \rightarrow \mu^- \mu^+) & = & 3.3662 \pm 0.0066 \%. 
\end{eqnarray}

The bound is depicted in Fig. \ref{fig_previous_bounds} on the top left as a slowly growing line (in magenta). Even after removing phase space suppression due to the lepton masses, $\Gamma(Z \rightarrow \tau^- \tau^+)$ still has some tension from the averaged value of leptonic decay width. If we specify our case as $U(1)_{L_\mu - L_\tau}$, it gives slightly stronger bound. However, in any case, the bounds from virtual corrections are much weaker than $\nu$-trident production bound.

\item \emph{Neutrino trident production}. ($\nu N \rightarrow \nu N \mu^+ \mu^-$). The neutrino-nucleon scattering experiments effectively provide the stringent constraint to the light gauge boson parameters which couple to the muon and the neutrino(s). The total cross section of $\nu$-trident production $\nu N \rightarrow \nu N \mu^+ \mu^-$ with $X$ boson, in the light $X$ boson limit ($m_X < m_\mu \ll \sqrt{s}$), is given by \cite{Altmannshofer:2014pba}
\begin{eqnarray}
\sigma^{\text{(SM+$X$)}} & = & \sigma^{\rm (SM)} + \sigma^{\rm (inter)} + \sigma^{(X)}, \nonumber \\
\sigma^{\rm(SM)} & \simeq & \frac{1}{2} (C_V^2 + C_A^2) \frac{2 G_F^2 \alpha s}{9\pi^2} \Bigl ( \ln \frac{s}{m_\mu^2} - \frac{19}{6} \Bigr ), \nonumber \\
\sigma^{\rm(inter)} & \simeq & \frac{G_F}{\sqrt{2}} \frac{g_X^2 C_V \alpha}{3\pi^2} \ln^2 \frac{s}{m_\mu^2}, \nonumber \\
\sigma^{(X)} & \simeq & \frac{1}{m_\mu^2} \frac{7 g_X^4 \alpha}{72\pi^2} \ln \frac{m_\mu^2}{m_X^2}.
\end{eqnarray} 

The CCFR experiment using a $\nu$-beam with $E_\nu \simeq 160$ GeV has obtained the result $\sigma_{\rm CCFR}/\sigma_{\rm SM} =0.82 \pm 0.28$ \cite{Mishra:1991bv}. The bound is depicted in Fig. \ref{fig_previous_bounds} by the purple line slightly above the $\pm 2 \sigma$ band of $(g-2)_\mu$.

\item \emph{Rare kaon decay at Beam-dump experiments}. Rare kaon decay at NA62 beam-dump experiment provides upper bound for muon-philic light bosons by rare kaon decay $K^+ \rightarrow \mu^+ \nu_\mu X (\rightarrow \nu \bar{\nu})$ for $m_X < 2 m_\mu$, and $K^+ \rightarrow \mu^+ \nu_\mu X (\rightarrow \mu^+ \mu^-)$ for $m_X \geq 2 m_\mu$ with a significant feature of some kinematic variables. Current bound comes from the $10^8$ charged kaons and it gives the upper bound as $g_X \lsim 10^{-2}$ \cite{Krnjaic:2019rsv} in the parameter range of our interests (also shown in Fig. \ref{fig_previous_bounds} by yellow line), although it is above the bound from neutrino trident experiment.\footnote{Recently, the future expected sensitivity from $10^{13}$ charged kaon and its rare decay such as $K^+ \rightarrow \mu^+ \nu_\mu X(\rightarrow \nu \bar{\nu}, \mu^- \mu^+)$ at NA62 experiment is explored in Ref. \cite{Krnjaic:2019rsv}. We show this result in Fig. \ref{fig_Belle2_mu_philic_sensitivity} (by yellow dashed line.)} \footnote{If one considers the kinetic mixing between $X$ boson and SM photon, it is also constrained by the channel $K^+, \pi^+ \rightarrow \mu^+ \nu_\mu e^- e^+$ \cite{Chiang:2016cyf} down to $\epsilon_{\gamma X} \sim \mathcal{O}(10^{-3} - 10^{-4}$).}

\item \emph{BaBar $4\mu$ channel search}. The BaBar experiment have explored \cite{TheBABAR:2016rlg} the muon-philic gauge boson by using the $4\mu$ channel ($e^- e^+ \rightarrow \mu^- \mu^+ X , \ X \rightarrow \mu^- \mu^+$), although the result is valid for the case $m_X > 2m_{\mu}$. This is depicted in Fig \ref{fig_previous_bounds} by the green colored (wiggly) region above $2m_\mu$.

\item \emph{Constraints from Big Bang Nucleosynthesis (BBN)}. A light $X$ boson coupled to neutrinos can directly enhance the number of relativistic degree of freedom in the BBN era for $m_X \lsim \mathcal{O}(1)$ MeV. Even in the heavier case $m_X \sim \mathcal{O}(1-10)$ MeV, the presence of muon-philic $X$ boson can affect the effective number of the light neutrino species $N_{\rm eff}$ by providing additional energies to $\nu_\mu, \ \bar{\nu}_\mu$ (and also $\nu_\tau, \ \bar{\nu}_\tau$ in $L_\mu - L_\tau$ case) from the decay process $X \rightarrow \nu_{\mu(\tau)} \bar{\nu}_{\mu(\tau)}$ after all SM neutrinos are decoupled from SM thermal bath at $T_{\nu,\text{dec}} \simeq 1.5$ MeV \cite{Kamada:2015era, Kamada:2018zxi, Escudero:2019gzq}. The deviation of the effective neutrino number $\Delta N_{\rm eff}$ comes from the difference between the tempreature $T'$ of the thermal bath of ($\nu_\mu, \bar{\nu}_\mu, \nu_\tau, \bar{\nu}_\tau, X$) and the temperature $T$ of the thermal bath of ($\nu_e, \bar{\nu}_e, \gamma$). This process is an analogy to the photon heating by the $e^- e^+ \rightarrow \gamma \gamma$ annihilation. Requiring $\Delta N_{\rm eff} < 0.7 \ (0.1)$, it disfavors the case $m_X \lsim 5.3 \ (10)$ MeV \cite{Kamada:2015era}\footnote{If one considers a non-negligible kinetic mixing between $X$ gauge boson and SM hypercharge gauge boson, the interaction between $X$ and $e^\pm$ also can affect $N_{\text{eff}}$ \cite{Kamada:2018zxi, Escudero:2019gzq}. In this case, the BBN bound can be slightly more stringent.}. The lower bound for $m_X$ corresponding to $\Delta N_{\rm eff} < 0.7$ is shown in Fig. \ref{fig_previous_bounds} as the orange dotted line. Because $X$ gauge boson can be in thermal equilibrium with other SM particles at early times as long as the coupling between $X$ and $\nu_{\mu,\tau}$ are $g_X \gsim 4 \times 10^{-9}$ \cite{Escudero:2019gzq}, this lower bounds on $m_X$ is valid in the range of our interest.

\end{itemize}

At low mass region ($m_X < 2 m_\mu$), $e^- e^+ \rightarrow \gamma X, \ X \rightarrow (\text{invisible})$ is the main channel of the minimal dark photon search \cite{Lees:2017lec}. The discovery potentials in the same channel $e^- e^+ \rightarrow \gamma X$ at Belle~II experiment also have been explored \cite{Kaneta:2016uyt, Araki:2017wyg}. 
 
However, the kinetic mixing between $X$ boson and SM $U(1)_Y$ is not determined, unless we assume that $\mu$- and $\tau$-lepton loops only contribute to the kinetic mixing $\epsilon_{\gamma X}$ which is the minimal mixing case. For instance, other heavy particle loops (from the particles with mass splitting $M_\psi - M_{\psi'}$) also can contribute to the mixing as $\epsilon_{\gamma X} \sim \frac{e g_X}{16\pi^2} \ln (M_{\psi}/M_{\psi'})$ and the total kinetic mixing depends on UV structure. For instance, one can modify the kinetic mixing with extra heavy vector-like leptons \cite{Chen:2017cic} or charged scalars \cite{Banerjee:2018mnw}. If one does not impose the kinetic mixing values between $U(1)_X$ and SM hypercharge gauge boson as $\epsilon_{\gamma X} \sim \mathcal{O}(10^{-4}$), the bound for purely light muon-philic force is not completely determined by low-energy $e^-e^+$-collider experiments up to now.
 
Similarly, other indirect bounds of muonic force for $m_X < 2m_\mu$, which comes from the electron-neutrino scattering process \cite{Bauer:2018onh} such as Borexino experiment \cite{Kaneta:2016uyt} and the white dwarf cooling \cite{Dreiner:2013tja}, also depend on the kinetic mixing between $X$ gauge boson and SM $U(1)_Y$ gauge boson, since these bounds assume $\nu_l$-$e^-$ scattering via $t$-channel with the mixing $\epsilon_{\gamma X}$.

Another advantage of considering the parameter region $m_X < 2m_\mu$ is to avoid stringent constraint from cosmic microwave background (CMB). In general, any symmetric population of dark matter (DM) particles which annihilate to particles in $s$-wave which inject electromagnetic energy can modify the CMB spectrum, so there is a stringent bound on this scenario. That would be the case, if $X$ boson becomes a portal to the dark sector. The upper bound for the annihilation cross section from CMB can be estimated by $\langle \sigma v \rangle /m_{\rm DM} \lsim 4.1 \times 10^{-28} \ {\rm cm^3 \ s^{-1} \ GeV^{-1}}$ \cite{Slatyer:2015jla}, which rules out the thermally produced DM lighter than 100 GeV. However, the CMB stringent constraint can be avoided, if $X$ decays only into invisible channels. In addition, since the kinetic mixing $\epsilon_{\gamma X}\sim \mathcal{O}(10^{-5})$ from $\mu-$ and $\tau-$loops is small enough, the process ${DM+DM}\to X+X \to X+\gamma$ could not give significant modification to CMB spectrum. Eventually, the bound from CMB can be satisfied in the parameter region $m_X < 2m_\mu$.

%%%%%%%%%%%%%%%%%%%%%%%%
%%%%%%%%%%%%%%%%%%%%%%%%

%%%%%%%%%%%%%%%%%%%%%%%%
\section{Expected sensitivity at Belle~II \label{sec:BELLE}} 
%%%%%%%%%%%%%%%%%%%%%%%%

\begin{figure}[h]
\centering
\includegraphics[width=0.65\textwidth]{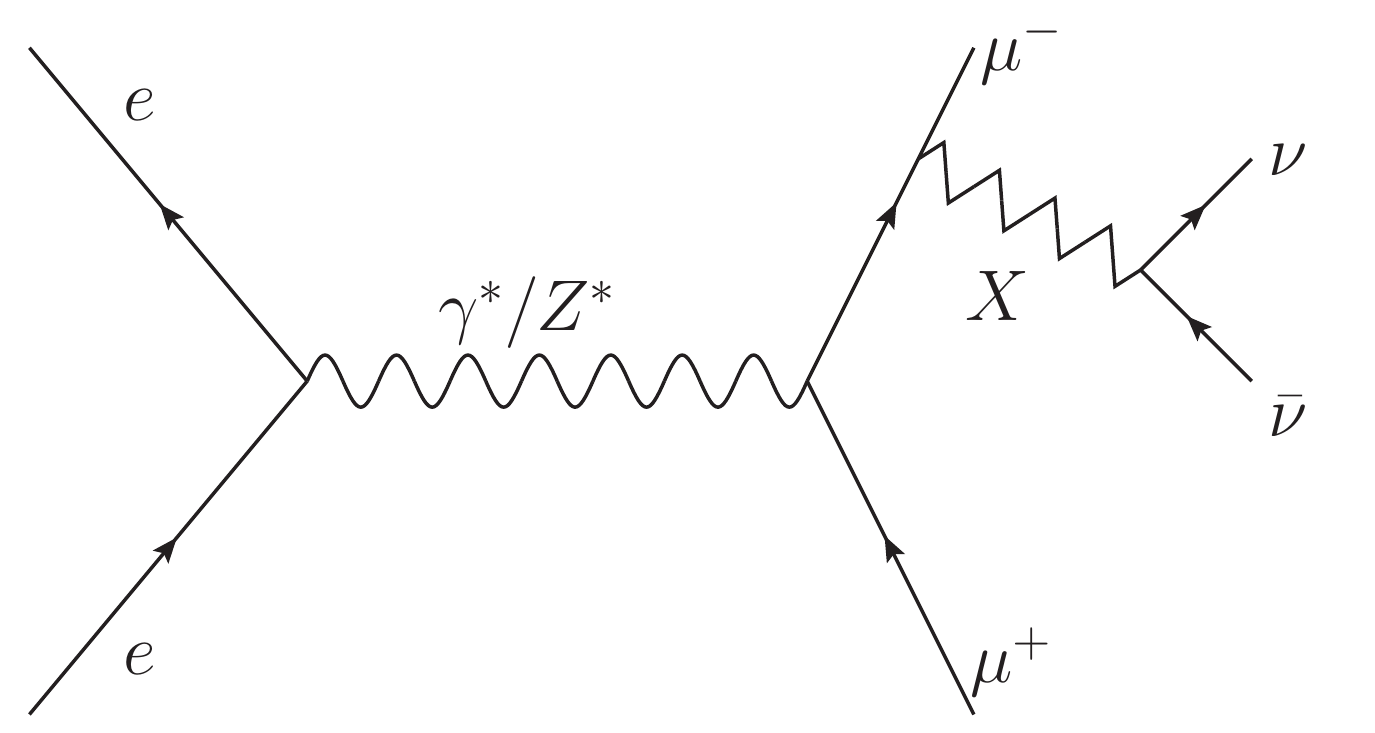}
\caption{The $X$ gauge boson production in $e^+e^-\to \mu^+\mu^- X$ channel.}
\label{fig:Feynman}
\end{figure}

The muon-philic gauge bosons are exclusively produced in muon-associated channels thus is less constrained compared with the model with universal couplings to fermions. In the Belle~II experiment, muons are pair-produced and $X$ boson can be radiated away from muon as in Fig.~\ref{fig:Feynman}. Finally, $X\to \nu\bar{\nu}$ and do not leave a detectable signal so that we regard it as an invisible particle (INV) and exploit appropriate kinematical variables such as missing transverse energy ($\slashed{E}_{T}$) and missing-mass-squared ($m_{\rm miss}^2$). Since the cross section of $e^+e^- \to \mu^+\mu^- X$ is proportional to $g_X^2 =4\pi \alpha_X \sim 8\pi^2 \Delta a_\mu^X$ so that we can almost directly check whether the $X$ boson would be responsible for the anomalous magnetic moment of muon from the measurement at Belle~II experiment. We provide some details about the expected sensitivity of muon-philic $X$ boson search in the $\mu^- \mu^+ +$INV channel at the Belle II experiment.
%%%%%%%%%%%%%%%%%%%%%%%%
%%%%%%%%%%%%%%%%%%%%%%%%

\subsection{Signal: $e^- e^+ \rightarrow \mu^- \mu^+ X, \ X \rightarrow \nu \bar{\nu}$}

\begin{figure}[h]
\centering
\includegraphics[width=0.48\textwidth]{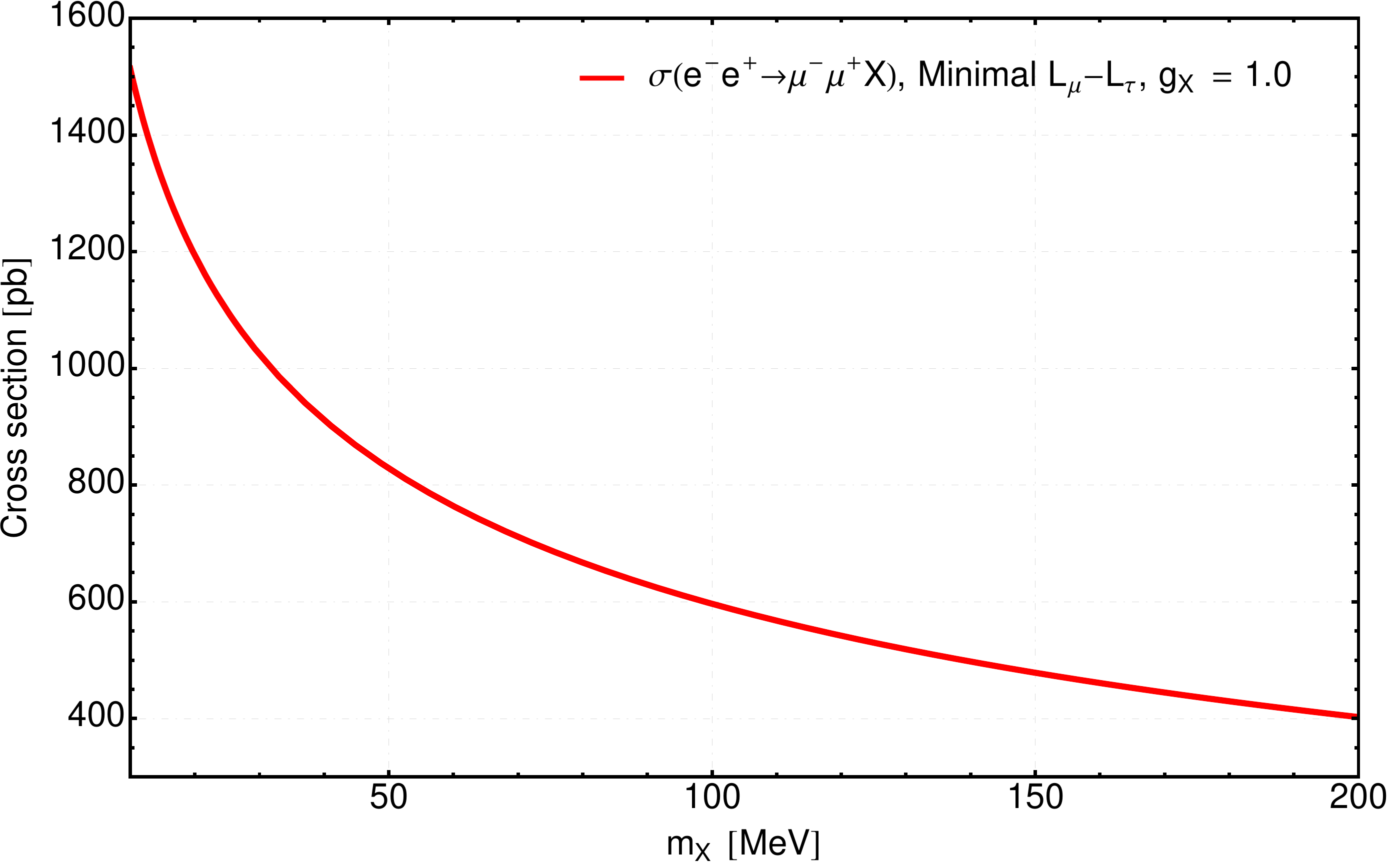} \,
\includegraphics[width=0.45\textwidth]{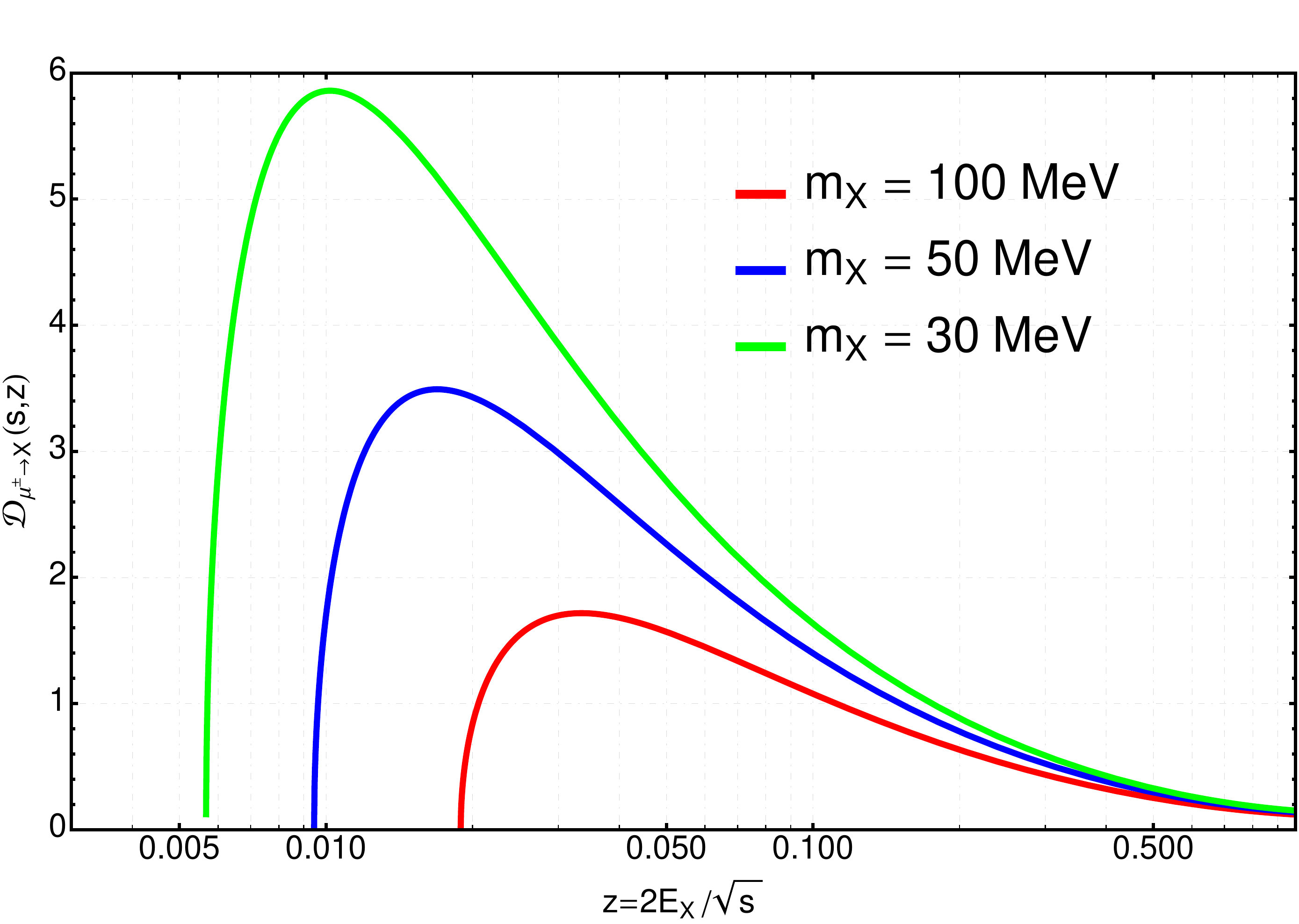}
\caption{(Left) The signal cross section $\sigma(e^- e^+ \rightarrow \mu^- \mu^+ X)$ without any kinematic cut, for $g_X = 1$. (Right) The splitting function $\mathcal{D}_{\mu^\pm \rightarrow X} (s, z)$ with $g_X = 1$ for the $X$ boson emission from muons.} \label{fig_prod_xsec_and_splitting_func} 
\end{figure}

The signal process is a muon-pair production with the real emission of a light $X$ boson as a final state radiation. Thus, most of $X$ bosons are very soft and collinear (along with muons' momenta). The signal cross section is \cite{Carone:1994aa}
\begin{eqnarray}
\sigma_{ee \rightarrow \mu \mu X} (s) & = & \sigma_{ee \rightarrow \mu \mu}^{(0)} (s) \cdot \frac{g_X^2}{8\pi^2} F_1 \Bigl ( \frac{m_X^2}{s} \Bigr )
\end{eqnarray}
where
\begin{eqnarray}
F_1 ( x ) & \equiv & (1+x)^2 \Bigl [ 3\ln x + (\ln x)^2 \Bigr ] + 5 (1-x^2) - 2 x \ln x \nonumber \\
& & \ - 2 (1+x)^2 \Bigl [ \ln (1+x) \ln x + \text{Li}_2 \Bigl ( \frac{1}{1+x} \Bigr ) - \text{Li}_2 \Bigl ( \frac{x}{1+x} \Bigr ) \Bigr ]
\end{eqnarray}
where $\sigma_{ee \rightarrow \mu \mu}^{(0)} (s) = 2 \pi \alpha^2 b (3-b^2)/(3s)$ is the cross section of muon pair production in the Born approximation with $b = 1 - \frac{4 m_\mu^2}{s}$. The cross section blows up as $m_X \rightarrow 0$ due to the infrared divergence as in usual final state radiation emission cases. (See Fig. \ref{fig_prod_xsec_and_splitting_func})

%%%%%%%%%%%%%%%%%%%%%%%%
%%%%%%%%%%%%%%%%%%%%%%%%

Including $X$ and muon masses, we utilize the splitting function of the $X$ emission in the process $\mu^\pm \rightarrow \mu^\pm + X$ for massive partons ($m_\mu, m_X \neq 0$) \cite{Ciafaloni:2001mu, Ciafaloni:2005fm, Ciafaloni:2010ti} as follows:
\begin{eqnarray}
\frac{d\sigma_{ee \rightarrow \mu \mu X} (s)}{dz}& = & \sigma_{ee \rightarrow \mu \mu}^{(0)} (s) \cdot 2 \mathcal{D}_{\mu^\pm \rightarrow X} (s, z)
\end{eqnarray}
where
\begin{eqnarray}
\mathcal{D}_{\mu^\pm \rightarrow X} (s, z) & = & \frac{g_X^2}{8\pi^2} \frac{1+ (1-z)^2}{z} \left [ \ln \frac{s z^2}{4 m_X^2} + 2 \ln \left ( 1 + \sqrt{1 - \frac{4 m_X^2}{s z^2} } \ \right ) \right ]
\end{eqnarray}
in the small mass limit $m_X \ll \sqrt{s}$ (See Fig. \ref{fig_prod_xsec_and_splitting_func}). Note that factor of 2 comes from $X$ boson emission by both $\mu^-$ and $\mu^+$. Here, $z \equiv E_X / (\sqrt{s}/2)$ is the energy fraction carried by the emitted $X$ boson, within kinematically allowed range
\begin{eqnarray}
\frac{2m_{X}}{\sqrt{s}} = z_{\min} \leq & z & \leq z_{\max} = 1 - \frac{2m_\mu}{\sqrt{s}},
\end{eqnarray}
and the total cross section is consistently given by integrating the spectral splitting function $\mathcal{D}_{\mu^\pm \rightarrow X} (s, z)$ as
\begin{eqnarray}
\sigma_{ee \rightarrow \mu \mu X}^{\rm total} (s) & = & \sigma_{ee \rightarrow \mu \mu}^{(0)} (s) \cdot 2 \int_{z_{\min}}^{z_{\max}} dz \ \mathcal{D}_{\mu^\pm \rightarrow X} (s, z) = \sigma_{ee \rightarrow \mu \mu}^{(0)} (s) \cdot \frac{g_X^2}{8\pi^2} F_1 \Bigl ( \frac{m_X^2}{s} \Bigr ) 
\end{eqnarray}

In principle, the signal ($\mu^- \mu^+ +$ INV) has a peak in the missing-mass-squared
\begin{eqnarray}
m_{\rm miss}^2 & = & (E_{\rm cm} - E_{\mu^-} - E_{\mu^+})^2 - (\vec{p}_{\mu^-} + \vec{p}_{\mu^+})^2
\end{eqnarray} 
around $m_{\rm miss}^2 \simeq m_X^2$. The decay width of $X$ boson is given by 
\begin{eqnarray}
\Gamma (X \rightarrow \nu_\mu \bar{\nu}_\mu, \nu_\tau \bar{\nu}_\tau) & = & \sum_{\nu = \nu_\mu, \nu_\tau} \frac{g_X^2}{24\pi} m_X \Bigl ( 1 + \frac{m_\nu^2}{m_X^2} \Bigr ) \sqrt{1 - 4 \frac{m_\nu^2}{m_X^2} } 
\end{eqnarray}
and this width is very small $(\Gamma_{X \rightarrow \nu \bar{\nu}} \sim g_X^2 m_X / 12 \pi \ll m_X)$ in the region of our interests ($g_X \lsim 10^{-3}$) and the narrow width approximation (NWA) is valid in our event analysis. In this case, we are sure that the produced $X$ bosons are on-shell, and spectral shape of $m_{\rm miss}^2$ will be very clear.\footnote{If the coupling of $X$ to dark sector is large as $g_D \sim \mathcal{O}(1)$ and a number of species of light ($2m_{\chi_i} \lsim m_X$) dark sector particles are coupled to $X$ boson ($N_\chi \gg 1$), then the width 
\begin{eqnarray} 
\Gamma_X^{\rm total} \simeq \Gamma(X \rightarrow \chi \bar{\chi}) & = & \sum_i \frac{g_D^2}{12\pi} m_X \Bigl ( 1 + \frac{m_{\chi_i}^2}{m_X^2} \Bigr ) \sqrt{1 - 4 \frac{m_{\chi_i}^2}{m_X^2} } \ \gsim \  \mathcal{O}(m_X)
\end{eqnarray}
for additional Dirac fermions $\chi_i$ in the dark sector coupled to $X$ gauge boson, for example. Thus, the finite width effect becomes significant in this case. However, for relatively small value of width $\Gamma_{X,\text{total}} \lsim m_X$, the production cross section $\sigma(e^- e^+ \rightarrow \mu^- \mu^+ X, X\rightarrow \chi \bar{\chi})$ is almost constant (even after the $\slashed{E}_T$ and $m_{\rm miss}^2$ cuts) because the narrow width approximation (NWA) is valid. Thus, our conclusion about the sensitivity of $g_X$ is indeed independent to the detail of the dark sector in most cases.
}

However, once the detector resolution is involved, the peak of missing mass becomes much broad with Gaussian smearing \cite{deFavereau:2013fsa}. The tracking resolution of muon momenta in the central drift chamber (CDC) detector is given as 
\begin{eqnarray}
\sigma_{p_{\mu^\pm}}/p_{\mu^\pm} & = & 0.0011 p_{\mu^\pm} \text{[GeV]} \oplus 0.0025/\beta
\end{eqnarray} 
at Belle II experiment, where $p_{\mu^\pm}$ is momentum of the muon track \cite{Adachi:2018qme}. We use $\sigma_{p_{\mu^\pm}}/p_{\mu^\pm} = 0.005$ in our event analysis at the detector level. For typical momentum of muons $p_{\mu^\pm} \simeq 3-5$ GeV, the momentum resolution is about $\sigma_{p_{\mu^\pm}} \simeq 15-25$ MeV. Thus, at the low $X$ boson mass region ($m_X \lsim 50$ MeV), it is hard to expect that the signal peak is distinguished from the backgrounds without additional kinematic cuts to remove relatively huge SM backgrounds.

\subsection{SM backgrounds and kinematic cuts}

\begin{figure}[h]
\centering
  {\includegraphics[width=0.4\textwidth]{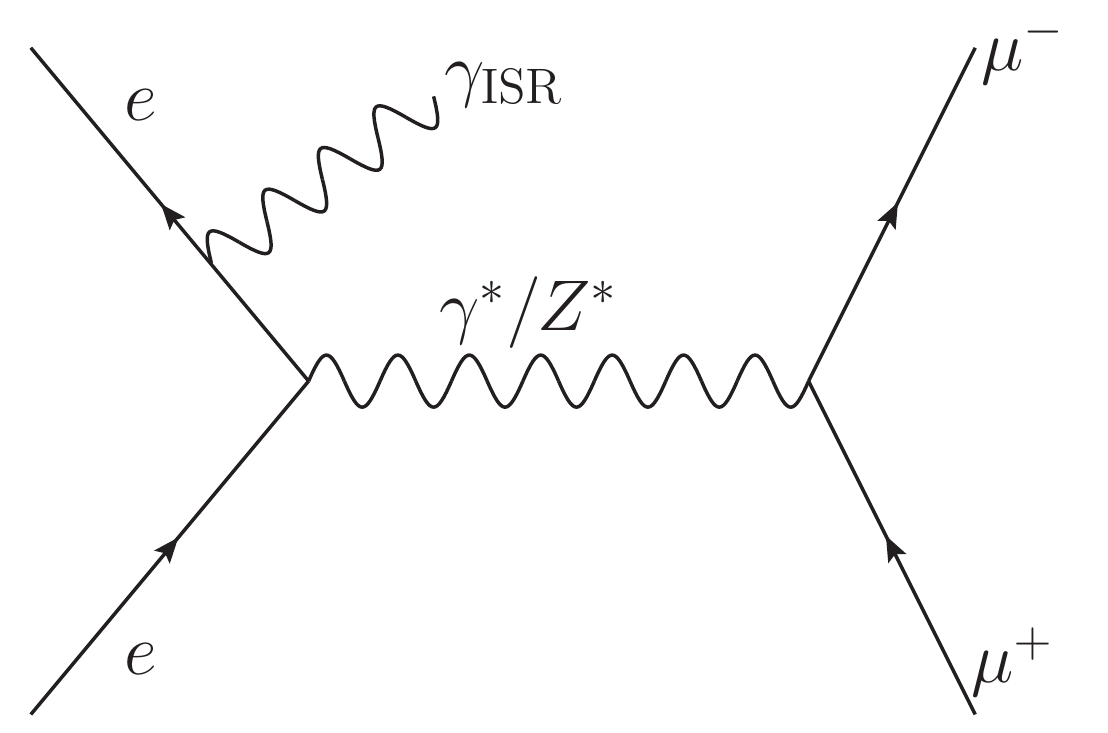}\label{fig_diag_bkgs_mumugamma_ISR}} \, \,
  {\includegraphics[width=0.4\textwidth]{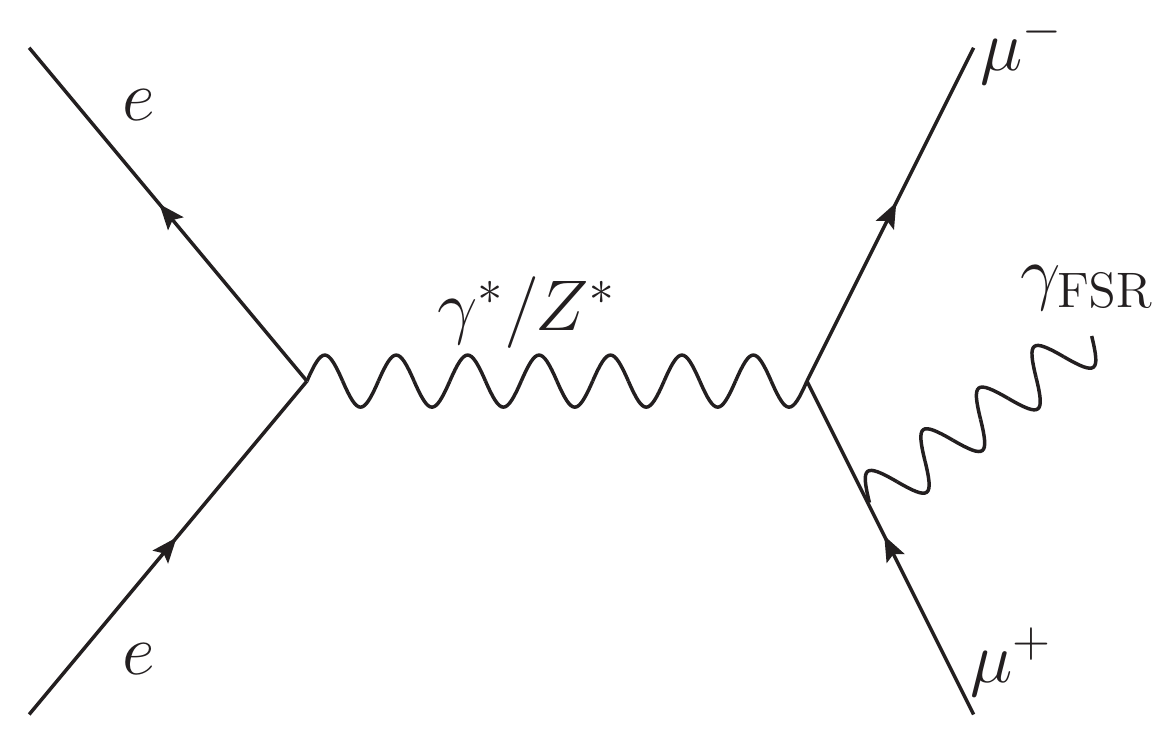}\label{fig_diag_bkgs_mumugamma_FSR}} \, \,
  {\includegraphics[width=0.4\textwidth]{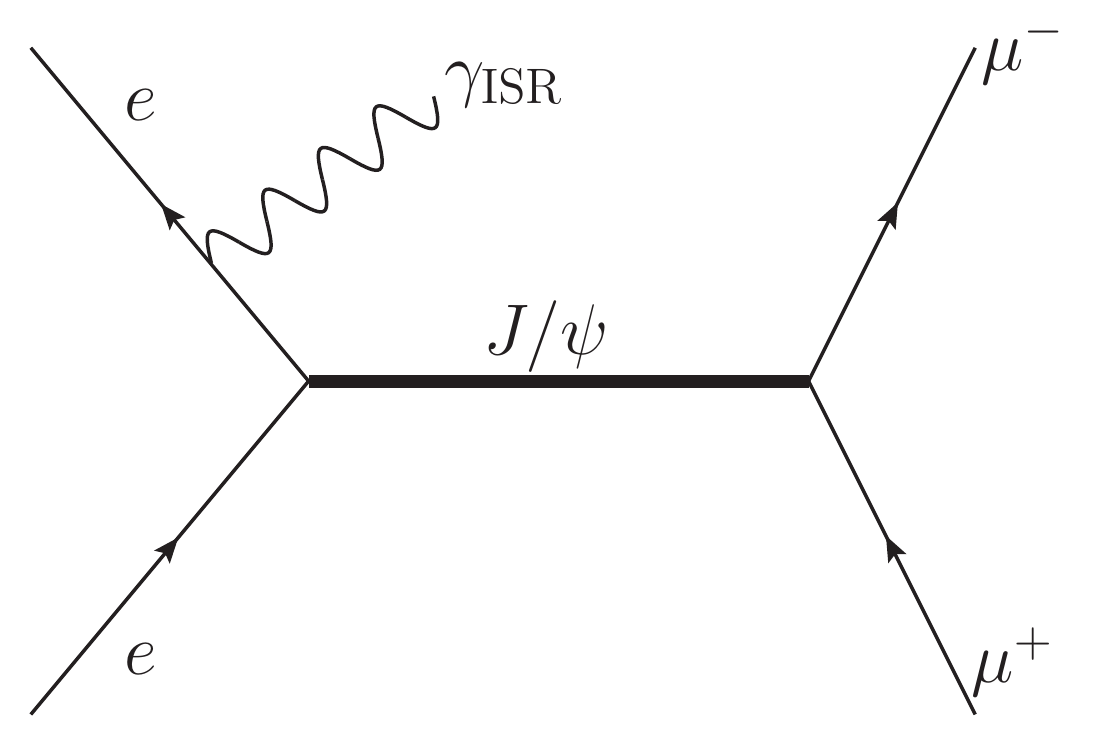}\label{fig_diag_bkgs_mumugamma_ISR_Jpsi}} \, \,
  {\includegraphics[width=0.45\textwidth]{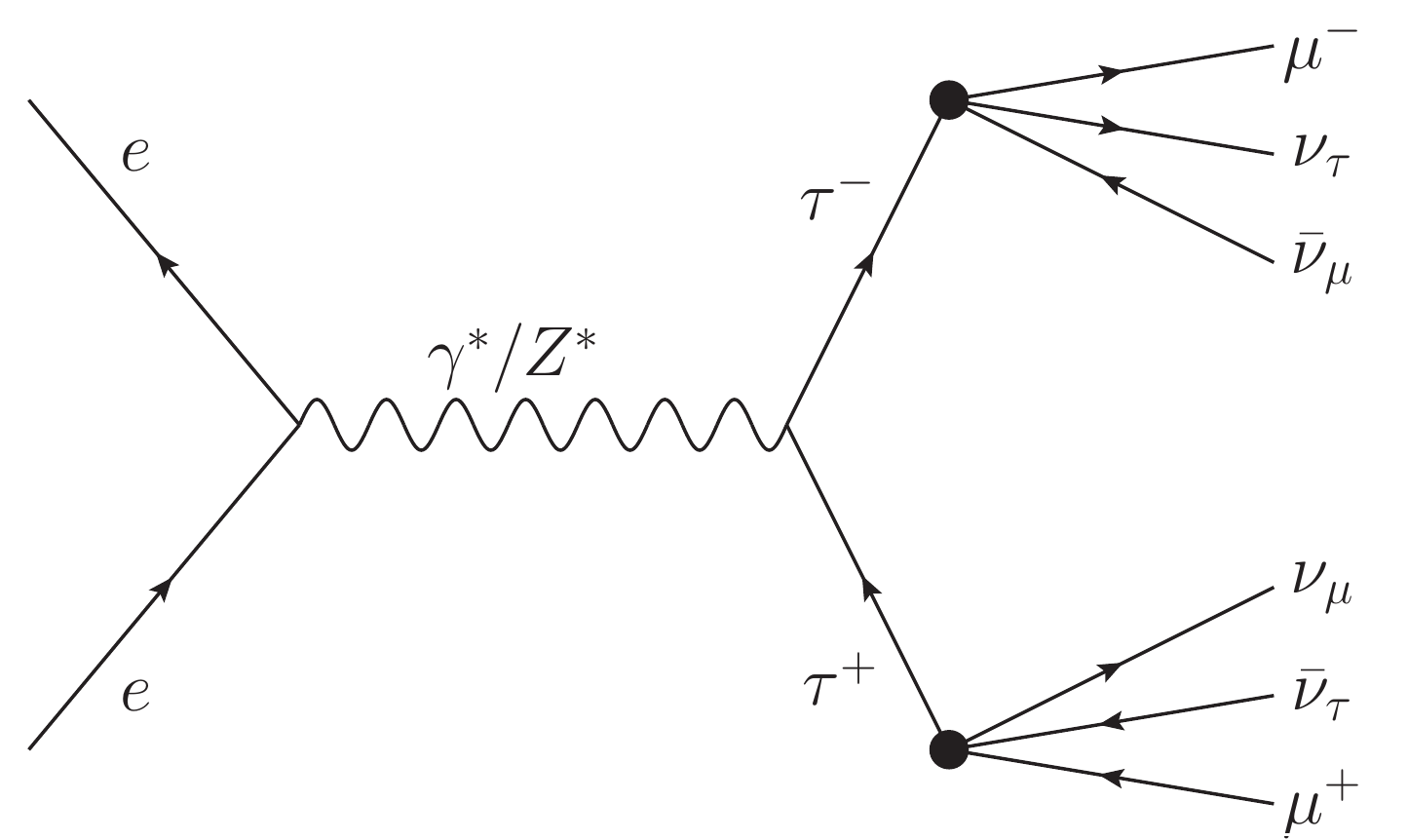}\label{fig_diag_bkgs_TauTau_muonically_decaying}} \, 
  {\includegraphics[width=0.35\textwidth]{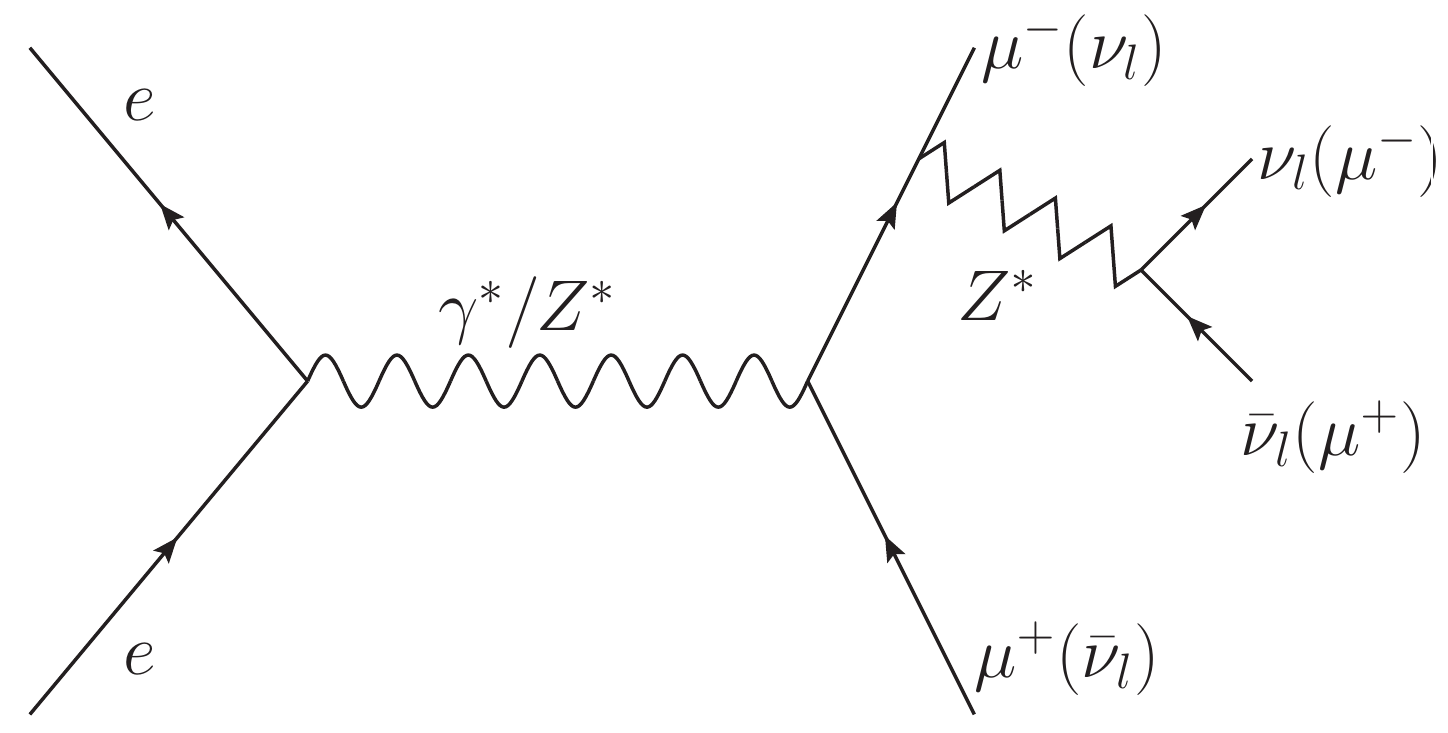}\label{fig_diag_bkgs_EW01}} \, \,
  {\includegraphics[width=0.24\textwidth]{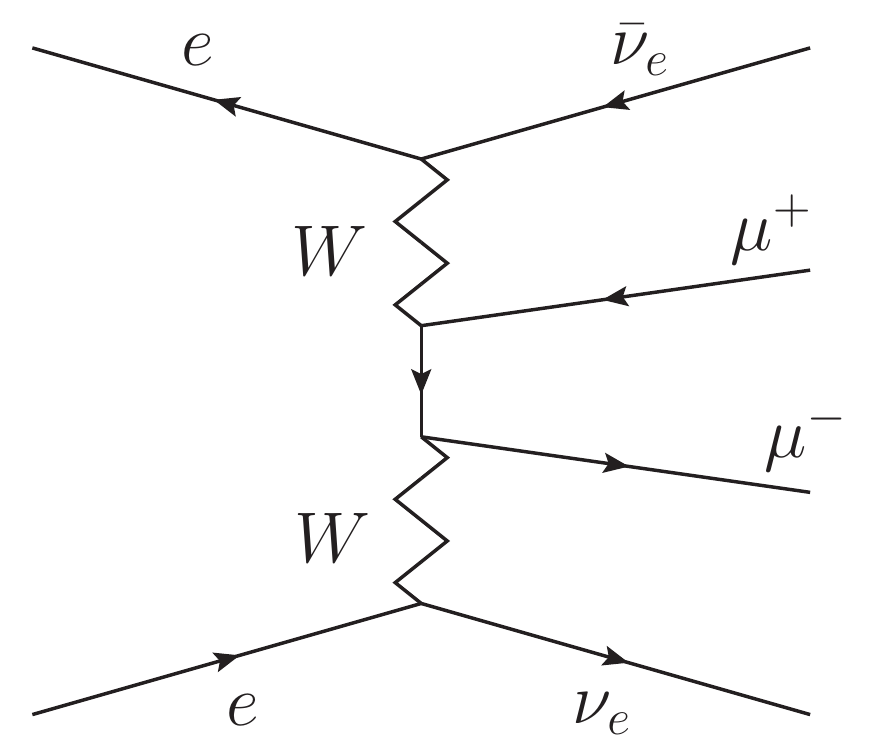}\label{fig_diag_bkgs_EW03}} \, \,   
  {\includegraphics[width=0.27\textwidth]{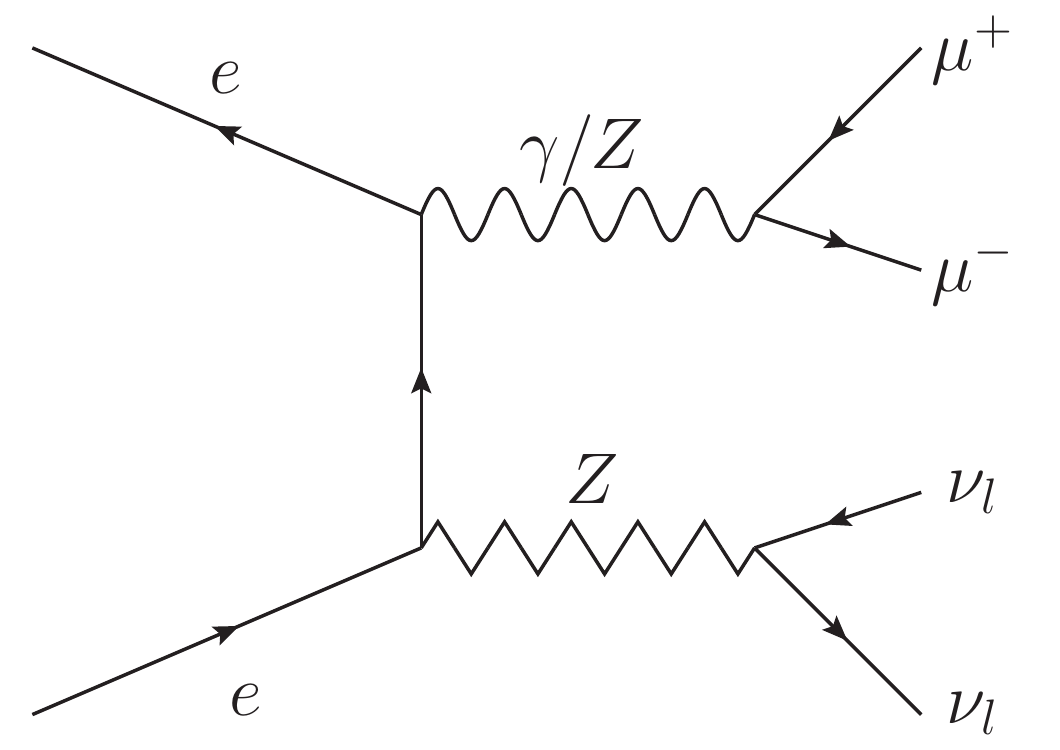}\label{fig_diag_bkgs_EW05}} \\ 
  {\includegraphics[width=0.35\textwidth]{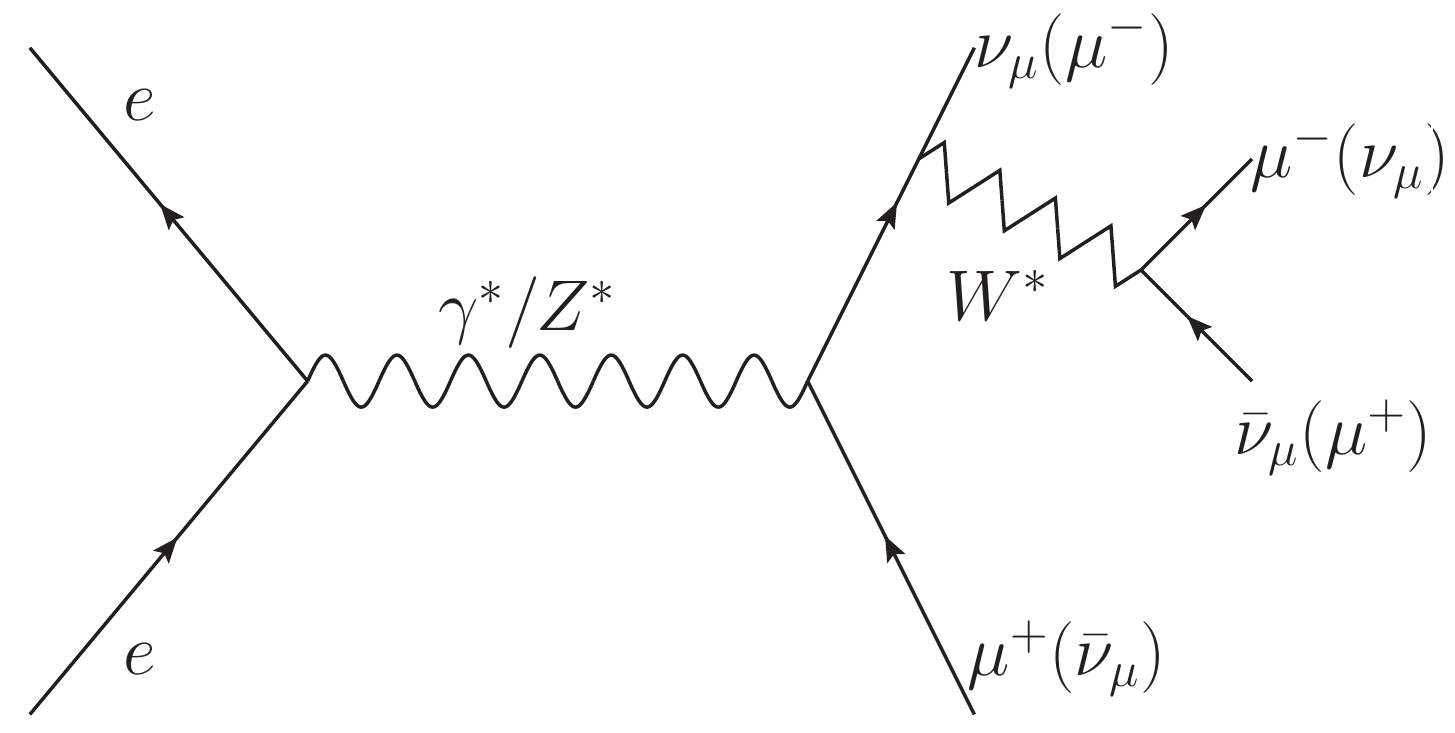}\label{fig_diag_bkgs_EW02}} \, \,
  {\includegraphics[width=0.24\textwidth]{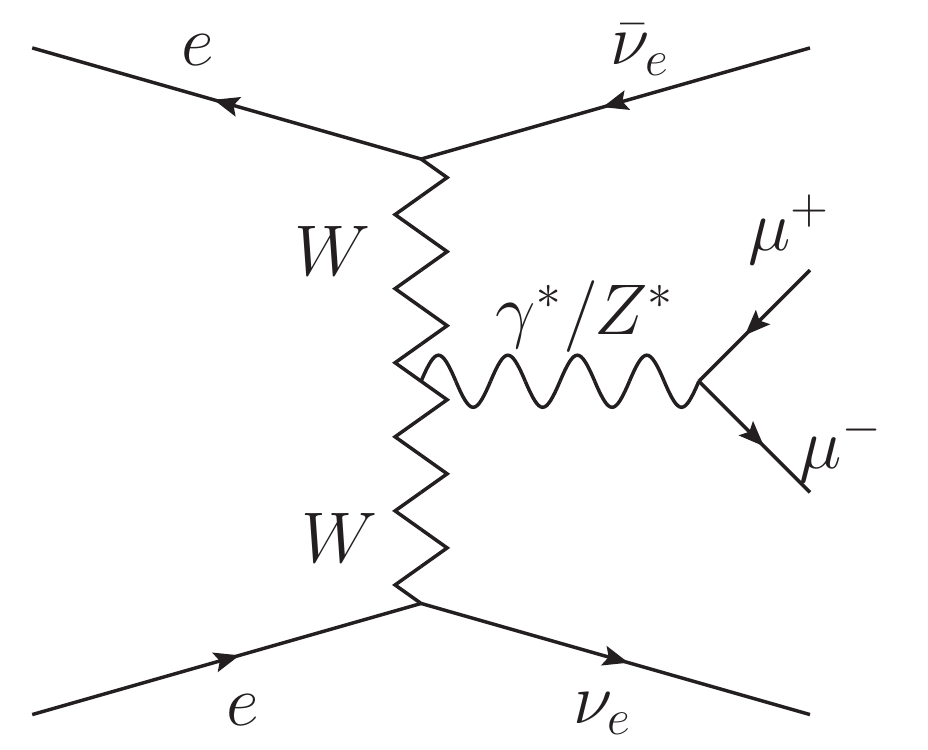}\label{fig_diag_bkgs_EW04}} \, \,
  {\includegraphics[width=0.27\textwidth]{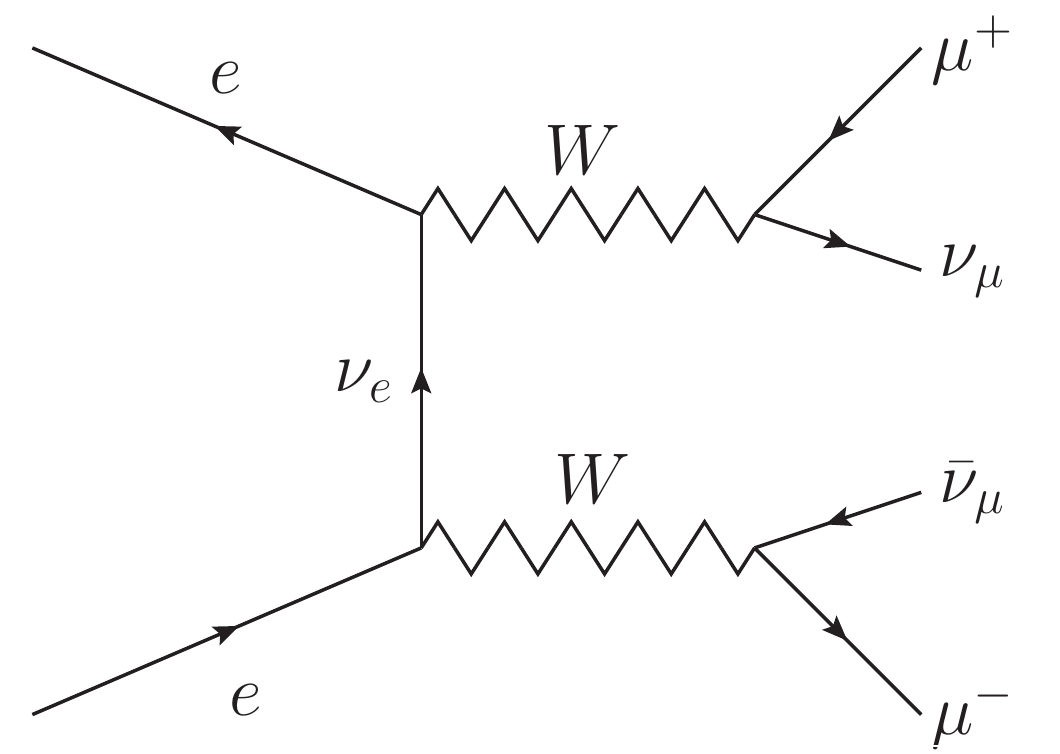}\label{fig_diag_bkgs_EW06}} 
  \caption{Main backgrounds from i) $\mu^- \mu^+ \gamma_{\rm ISR,FSR}$ (top), ii) $\mu^- \mu^+ \gamma_{\rm ISR}$ via $J/\psi$ meson production (middle left), iii) $\tau^- \tau^+ (\rightarrow \mu^- \mu^+ \nu_\mu \nu_\tau \bar{\nu}_\mu \bar{\nu}_\tau)$ (middle right), and iv) $W$- and $Z$- involved process (bottom).}
 \label{Backgrounds}
\end{figure}

The main $\mu^- \mu^+ + \slashed{E}_T$ backgrounds are as follows:
\begin{itemize}
\item $e^- e^+ \rightarrow \mu^- \mu^+ (\gamma_{\rm ISR, FSR})$

\item $e^- e^+ \rightarrow \tau^- \tau^+ \rightarrow \mu^- \mu^+ \nu_\mu \bar{\nu}_\mu \nu_\tau \bar{\nu}_\tau$

\item $e^- e^+ \rightarrow \mu^- \mu^+ \nu_l \bar{\nu}_l$ by off-shell $W$ and $Z$

\end{itemize}
and the diagrams for each background process are shown in Fig. \ref{Backgrounds}.

Most dominant background process is $\mu^- \mu^+ \gamma$, which has typically $\mathcal{O}(100)$ pb of the production cross section, although all of them actually can be removed using kinematic cuts. To remove $\mu^- \mu^+ \gamma_{\rm ISR}$ and $\mu^- \mu^+ \gamma_{\rm FSR}$ backgrounds, we reject all events with $\slashed{E}_T < 1.67$ GeV or with the photon energy in the center-of-mass frame $E_\gamma > 1.0$ GeV where the electromagnetic calorimeter (ECL) has high efficiency \cite{Adachi:2018qme}. This kinematic cut removes most of the $\mu^- \mu^+ \gamma_{\rm ISR,FSR}$ backgrounds. One notices that, at the center-of-mass energy $\sqrt{s} = 10.58$ GeV, the resonant production of $D$ and $B$ mesons are not negligible. Indeed, $J/\psi$ meson can be produced with a photon and decay into $\mu^- \mu^+$ (or $\tau^- \tau^+$), and its contribution to total $\mu^- \mu^+ \gamma$ production cross section is $\sim 0.12$ \% \cite{Aubert:2003sv, Banerjee:2007is}. However, due to its small $\slashed{E}_T$, most of the $J/\psi$ background events are removed by requiring $\slashed{E}_T > 1.67$~GeV.

For muonically decaying tau-pairs $\tau^- \tau^+ (\rightarrow \mu^- \mu^+ \nu_\mu \bar{\nu}_\mu \nu_\tau \bar{\nu}_\tau)$, the cross section is 
\begin{eqnarray}
\sigma (e^- e^+ \rightarrow \tau^- \tau^+, \ \tau^\pm \rightarrow \mu^\pm \nu_{\tau (\mu)} \bar{\nu}_{\mu (\tau)}) \approx 27.79 \text{ pb}
\end{eqnarray}
with the collision energy $\sqrt{s} = 10.58$ GeV~\cite{Banerjee:2007is}, and they contribute as a significant background. Although the final state ($\mu^- \mu^+ + $INV) is the same as the signal mode, its energy spectrum is completely different. The muons come from the decay of taus, and the muon energies at the muon pair center-of-mass frame have broad continuum distributions. In the center-of-mass frame of the electron-positron collision, the differential cross section of the (muonically decaying) tau pair production \cite{Scheck:1977yg, Ackerstaff:1998yk} is given by
\begin{eqnarray}
\frac{1}{\sigma} \frac{d^2 \sigma (x, \ \cos \theta, \ P_e)}{dx \ d \cos \theta} & = & f(x) - P_\tau (\cos \theta, \ P_e) \cdot g(x), 
\end{eqnarray}
where $x = E_\mu / E_\tau$ and $E_\tau = \sqrt{s}/2$. The distribution is given by
\begin{eqnarray}
f(x) & = & \Bigl ( 2 - 6 x^2 + 4 x^3 \Bigr ) + \rho_\mu \cdot \frac{4}{9} \Bigl (-1 + 9x^2 - 8 x^3 \Bigr ), \nonumber \\
g(x) & = & \xi_\mu \cdot \Bigl [ \Bigl ( - \frac{2}{3} + 4 x - 6 x^2 + \frac{8}{3} x^3 \Bigr ) + \delta_\mu \cdot \frac{4}{9} \Bigl ( 1 - 12x + 27 x^2 - 16 x^3 \Bigr ) \Bigr ], \nonumber \\
P_\tau ( \cos \theta, \ P_e) & = & - \frac{A_\tau + 2 \frac{A_e - P_e}{1 - A_e P_e} \frac{\cos \theta}{1+ \cos^2 \theta} }{1 + 2 A_\tau \frac{A_e - P_e}{1 - A_e P_e} \frac{\cos \theta}{1+ \cos^2 \theta} } 
\end{eqnarray}
where $A_l = \frac{2 \hat{g}_v^l / \hat{g}_a^l}{1 + (\hat{g}_v^l / \hat{g}_a^l)^2}$. Here, $\hat{g}_v^l$ and $\hat{g}_a^l$ is the vector and axial-vector couplings to the charged leptons. We use the Mitchel parameters $\rho_\mu = \frac{3}{4}$, $\xi_\mu = 1$, $\delta_\mu = \frac{3}{4}$ as the prediction in the Standard Model \cite{Ackerstaff:1998yk}. The anisotropic contribution is negligible because off-shell photon (not $Z$) is dominant channel for $\sqrt{s} \ll m_Z$ and initial electron and positron beams are not polarized. We use {\tt TauDecay} \cite{Hagiwara:2012vz} library to make {\tt FeynRules} \cite{Alloul:2013bka} model file which allows to perform $\tau$ decays with polarization. Most events in this background are in the region $m_{\rm miss}^2 \gsim (0.6~{\rm GeV})^2$, which is beyond the region of our interest ($m_X < 2 m_\mu = 0.211$ GeV). If one imposes the condition $\slashed{E}_T > 1.67$ GeV, the remaining $m_{\rm miss}^2$ values become even larger. Thus, we can safely ignore tau-pair background after $m_{\rm miss}^2$ cuts.

There are also off-shell $W$ and $Z$ involved process ($e^- e^+ \rightarrow \mu^- \mu^+ \nu \bar{\nu}$). The cross section is $\sim 7 \times 10^{-2}$ fb. However, it is 4-body production channel and highly off-shell, so after $\slashed{E}_T$ and $m_{\rm miss}^2$ cuts, no background events remain, even at the integrated luminosity of 50 ab$^{-1}$. 

%%%%%%%%%%%%%%%%%%%%%%%%
%%%%%%%%%%%%%%%%%%%%%%%%

\subsection{Event Analysis}

We use {\tt MadGraph5\_aMC@NLO} \cite{Alwall:2011uj} for background and signal event generation. We use our own {\tt FeynRules} \cite{Alloul:2013bka} model file for $X$ gauge boson coupled to muon (and neutrino), to generate $\mu^- \mu^+ X$ signal events. Event analyses have been performed for the following sets of Monte Carlo events ($5 \times 10^5$ events for each set):
\begin{itemize}
\item $e^- e^+ \rightarrow \mu^- \mu^+ (\gamma)$ \emph{background with} $| \eta_\gamma^* | > 1.94$\footnote{In this case, photons are highly collinear with beam axis, and just go through the beam pipe and only muon pair with some small $\slashed{E}_T$ in the final state.}
\item $e^- e^+ \rightarrow \mu^- \mu^+ (\gamma)$ \emph{background with $| \eta_\gamma^* | < 1.94$}
\item $e^- e^+ \rightarrow \tau^- \tau^+ (\rightarrow \mu^- \mu^+ \nu_\mu \nu_\tau \bar{\nu}_\mu \bar{\nu}_\tau)$ \emph{background}
\item \emph{off-shell} $W^*/Z^*$ \emph{involved} $e^- e^+ \rightarrow \mu^- \mu^+ \nu \bar{\nu}$ \emph{background}
\item $e^- e^+ \rightarrow \mu^- \mu^+ X$ \emph{signal}
\end{itemize}
where $\eta_\gamma^*$ is the photon rapidity in the center-of-mass frame and the muon rapidity in the center-of-mass frame $\eta_{\mu^\pm}^*$ is given in the range $-1.60 < \eta_{\mu^\pm}^* < 1.21$ for all events. All rapidity cuts are considered in the center-of-mass frame so that all muons are within both CDC ($17.0^\circ < \theta_{\mu^\pm}^{\rm lab.} < 150.0^\circ$) and $K_L$ and muon detector (KLM) ($25.0^\circ < \theta_{\mu^\pm}^{\rm lab.} < 155.0^\circ$) angle coverages and all photons are within ECL ($12.4^\circ < \theta_\gamma^{\rm lab.} < 155.1^\circ$) angle coverage \cite{Adachi:2018qme} after Lorentz boost with $\beta_{\rm Belle} = \frac{E_{e^-} - E_{e^+}}{E_{e^-} + E_{e^+}}$ is performed, where $E_{e^-} = 7.0$ GeV and $E_{e^+} = 4.0$ GeV. 

 The most dominant background source is the process $e^+ e^- \to \mu^+ \mu^- \gamma$ in which the photon is not detected. Typically, for Belle~II, the inefficiency is $1-\epsilon_\gamma \simeq 0.05$. It is mainly due to the small gaps between barrel and endcap regions ($31.4^\circ < \theta_\gamma^{\rm lab} < 32.2^\circ$ and $128.7^\circ < \theta_\gamma^{\rm lab} < 130.7^\circ$), a $1-1.5~{\rm mm}$ gap at $\theta^{\rm lab} = 90^\circ$ owing to the mechanical structure of the Belle~II ECL, and $0.2~{\rm mm}$ gaps between the crystals in the ECL endcap region. 

Most of this inefficiencies are removed by requiring the direction of missing 3-momentum (in this case, the 3-momentum of the unobserved photon) to be within the ECL barrel region.  The inefficiency $1-\epsilon_\gamma$ is then reduced to $\sim 3\times 10^{-6}$ \cite{Kou:2018nap} which comes from intrinsic probability of missing the photon detection inside the ECL crystals.
 
 For Belle~II, the KLM detector can also be used to detect photons. By combining the ECL and KLM together for photon detection, the inefficiency is suppressed down to $1-\epsilon_\gamma = 10^{-6}$. In fact, it can provide an improved sensitivity limit on the ``single-photon" search at Belle~II ($e^- e^+ \rightarrow \gamma X$)  down to $\epsilon_{\gamma X} \sim 3 \times 10^{-4}$ \cite{Kou:2018nap}. Therefore, in this paper, we set the conservative (aggressive) nominal value of photon inefficiency $1-\epsilon_\gamma = 10^{-5} \ (10^{-6})$. In addition, imposing $\slashed{E}_T$ and $m_{\rm miss}^2$ cuts and muon detection efficiency for these background events, the expected $\mu^- \mu^+ (\gamma)$ event number is $\sim 179.5 \left ( \frac{1-\epsilon_\gamma}{10^{-5}} \right ) \left ( \frac{\int \mathcal{L} \ dt}{1 \text{ ab}^{-1}} \right )$. In this study, the uncertainty in $1-\epsilon_\gamma$ becomes the dominant source of the systematic uncertainties.  The other sources such as in the selection of two muons and other kinematic requirements, in comparison, contribute much less to the total systematic uncertainty.

\begin{figure}[h]
\centering
\subfloat[$\mu^- \mu^+ \gamma$ with $|\eta_\gamma^*|>1.94$]
{\includegraphics[width=0.32\textwidth]{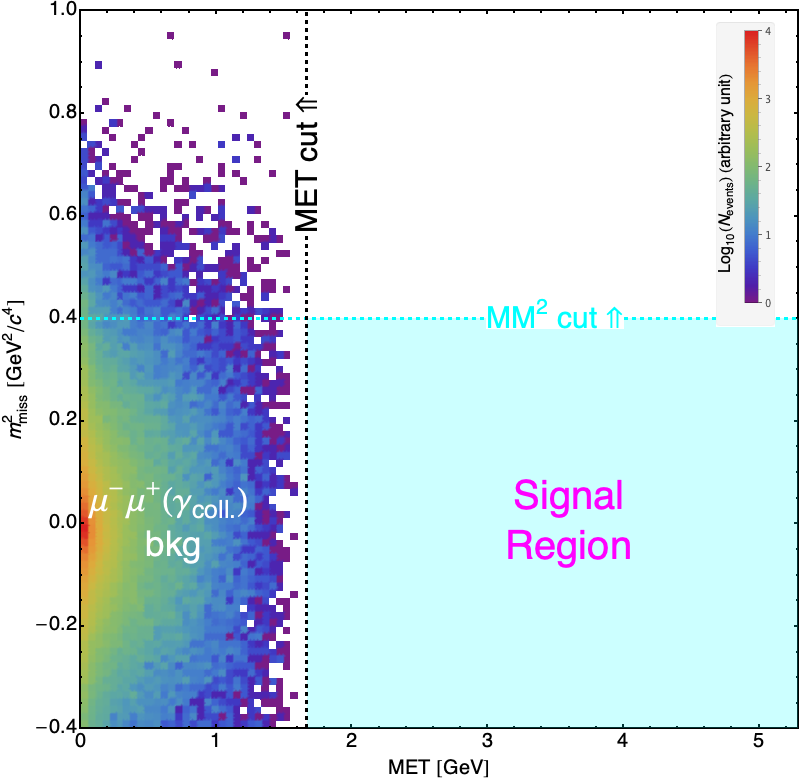}\label{fig_2Dplot_METMM2_bkg01_mumugamma_collinear}} \, 
\subfloat[$\mu^- \mu^+ \gamma$ with $|\eta_\gamma^*|<1.94$]
{\includegraphics[width=0.32\textwidth]{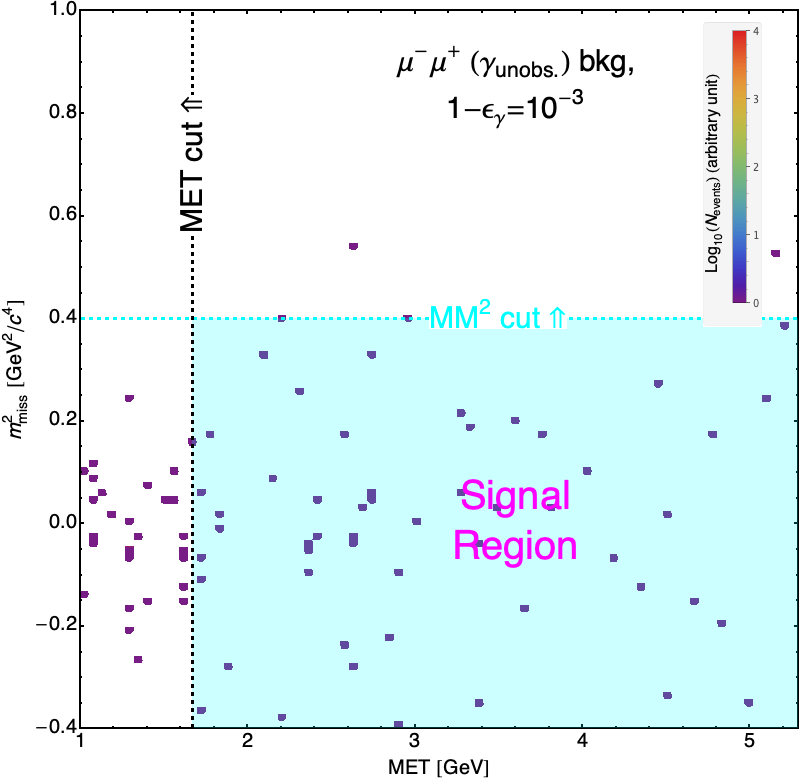}\label{fig_2Dplot_METMM2_bkg02_mumugamma_unobs_eff_1e-3}} \, 
\subfloat[$\tau^- \tau^+(\rightarrow \mu^- \mu^+ \nu \nu \bar{\nu} \bar{\nu})$]
{\includegraphics[width=0.307\textwidth]{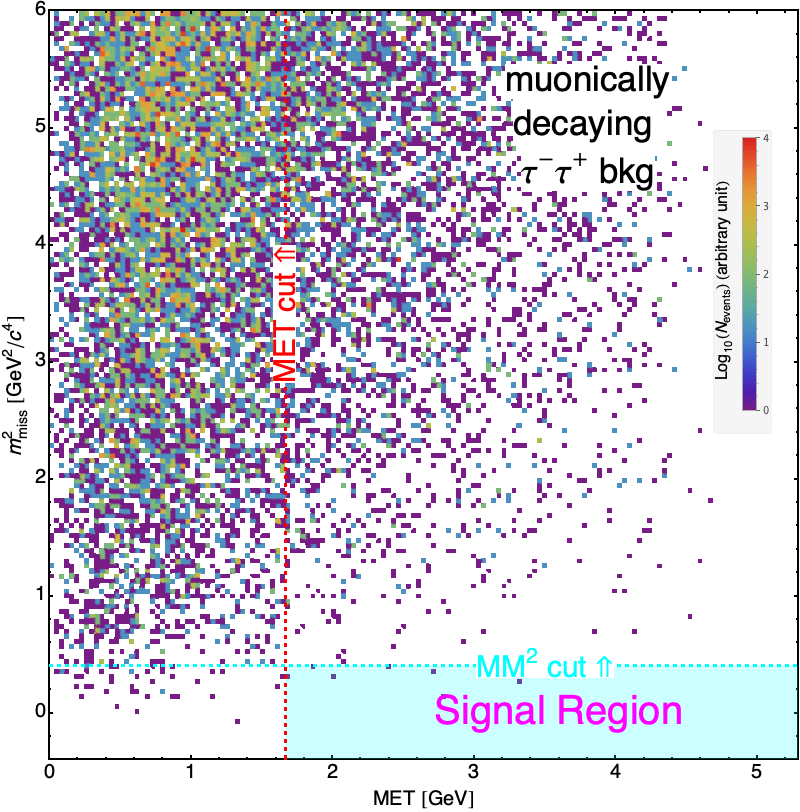}\label{fig_2Dplot_METMM2_bkg03_muonically_decaying_tautau}} 

\caption{Distributions of SM background events $\mu^-\mu^+ (\gamma)$, $\tau^- \tau^+(\rightarrow \mu^- \mu^+ \nu \nu \bar{\nu} \bar{\nu})$ on the plane of ($\slashed{E}_T$, $m_{\rm miss}^2$) space. In the case of $\mu^-\mu^+ +$ (unobserved) $\gamma$ background with $|\eta_\gamma^*|<1.94$, we show the case of photon detection inefficiency $1-\epsilon_\gamma = 10^{-3}$ for demonstration.} 
\label{METandMM2_SM_bkg}
\end{figure}

\begin{figure}[h]
\centering
{\includegraphics[width=0.47\textwidth]{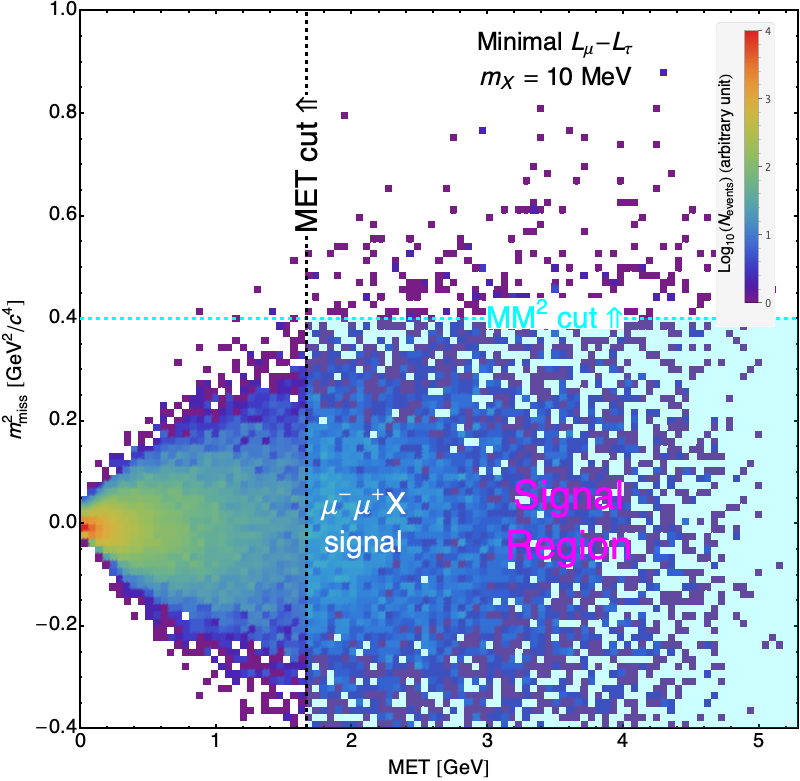}\label{fig_2Dplot_METMM2_sig01_01_minimal_LmuLtau_mX_10MeV}} \,
{\includegraphics[width=0.47\textwidth]{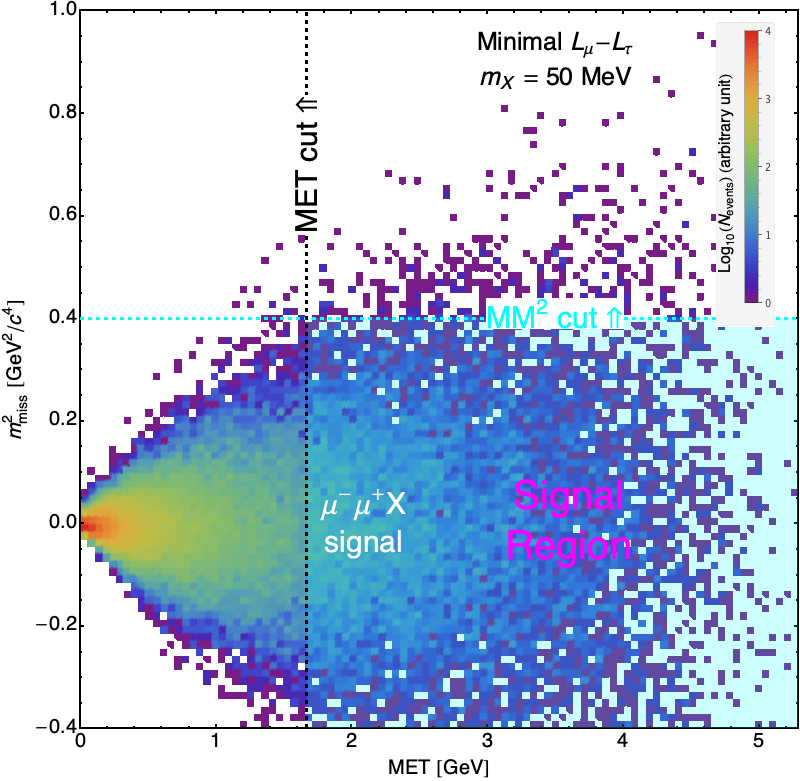}\label{fig_2Dplot_METMM2_sig01_02_minimal_LmuLtau_mX_50MeV}} \\
{\includegraphics[width=0.47\textwidth]{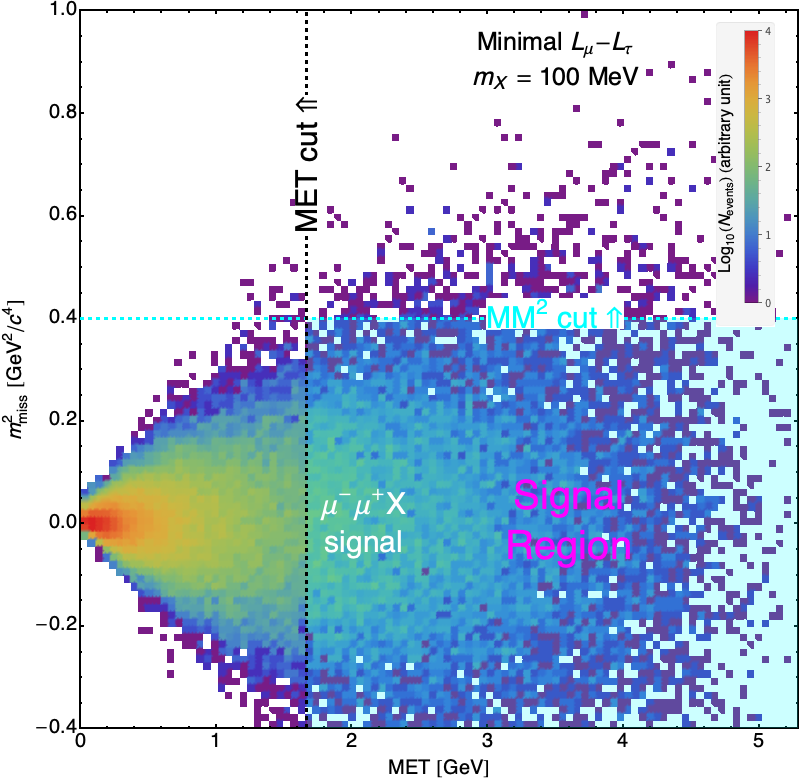}\label{fig_2Dplot_METMM2_sig01_03_minimal_LmuLtau_mX_100MeV}} \,
{\includegraphics[width=0.47\textwidth]{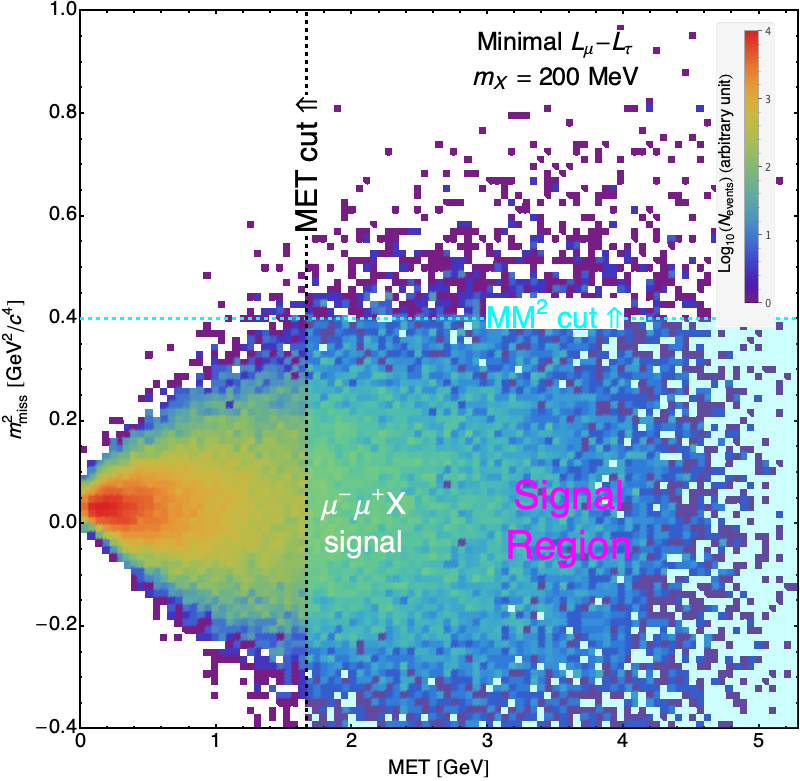}\label{fig_2Dplot_METMM2_sig01_04_minimal_LmuLtau_mX_200MeV}} 
\caption{Distributions of signal events $\mu^-\mu^+ X$ on the plane of ($\slashed{E}_T$, $m_{\rm miss}^2$) space for $m_X = 10, 50, 100, 200$ MeV.} 
\label{METandMM2_minimal_LmuLtau_sig}
\end{figure}

As we mentioned in the previous section, $\slashed{E}_T$, $E_\gamma$ and $m_{\rm miss}^2$ cuts are used to remove all background events. Comparison for signals and backgrounds under these kinematic variables are shown in Fig. \ref{fig_1Dplot_MET_cuts} for $\slashed{E}_T$ and Fig. \ref{fig_1Dplot_MissingMass_and_MET_cuts} for $m_{\rm miss}^2$. Also, we show correlations between $\slashed{E}_T$ and $m_{\rm miss}^2$ in Fig. \ref{METandMM2_SM_bkg} and Fig. \ref{METandMM2_minimal_LmuLtau_sig} for backgrounds and signals, respectively.

%%%%%%%%%%%%%%%%%%%%%%%%
\begin{figure}[h]
\centering
 {\includegraphics[width=0.7\textwidth]{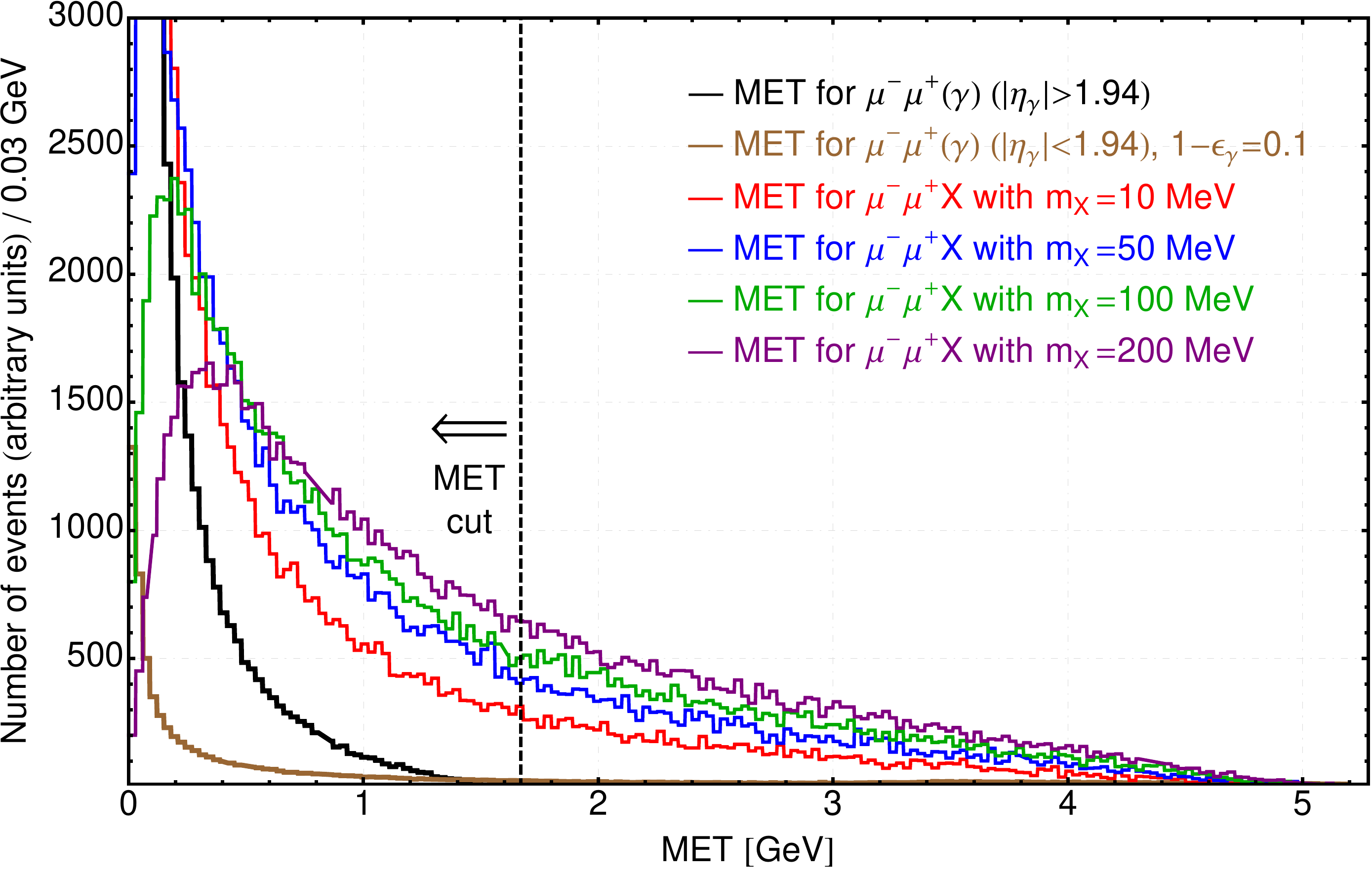}\label{fig_1Dplot_MET_cuts}} \, \, 
  \caption{$\slashed{E}_T$ distribution for $\mu^- \mu^+ \gamma$ backgrounds and $\mu^-\mu^+ X$ signal events. We only choose events with $\slashed{E}_T > 1.67$ GeV. We show $\mu^- \mu^+$ + unobserved $\gamma$ ($|\eta_\gamma|<1.94$) background events with photon detection inefficiency $1-\epsilon_\gamma = 10^{-1}$ for demonstration. The $\slashed{E}_T$ cut is shown as vertical dotted line. Each event set contains $1 \times 10^5$ events. \label{fig_1Dplot_MET_cuts}}
\end{figure}
%%%%%%%%%%%%%%%%%%%%%%%%

%%%%%%%%%%%%%%%%%%%%%%%%
\begin{figure}[h]
\centering
{\includegraphics[width=0.7\textwidth]{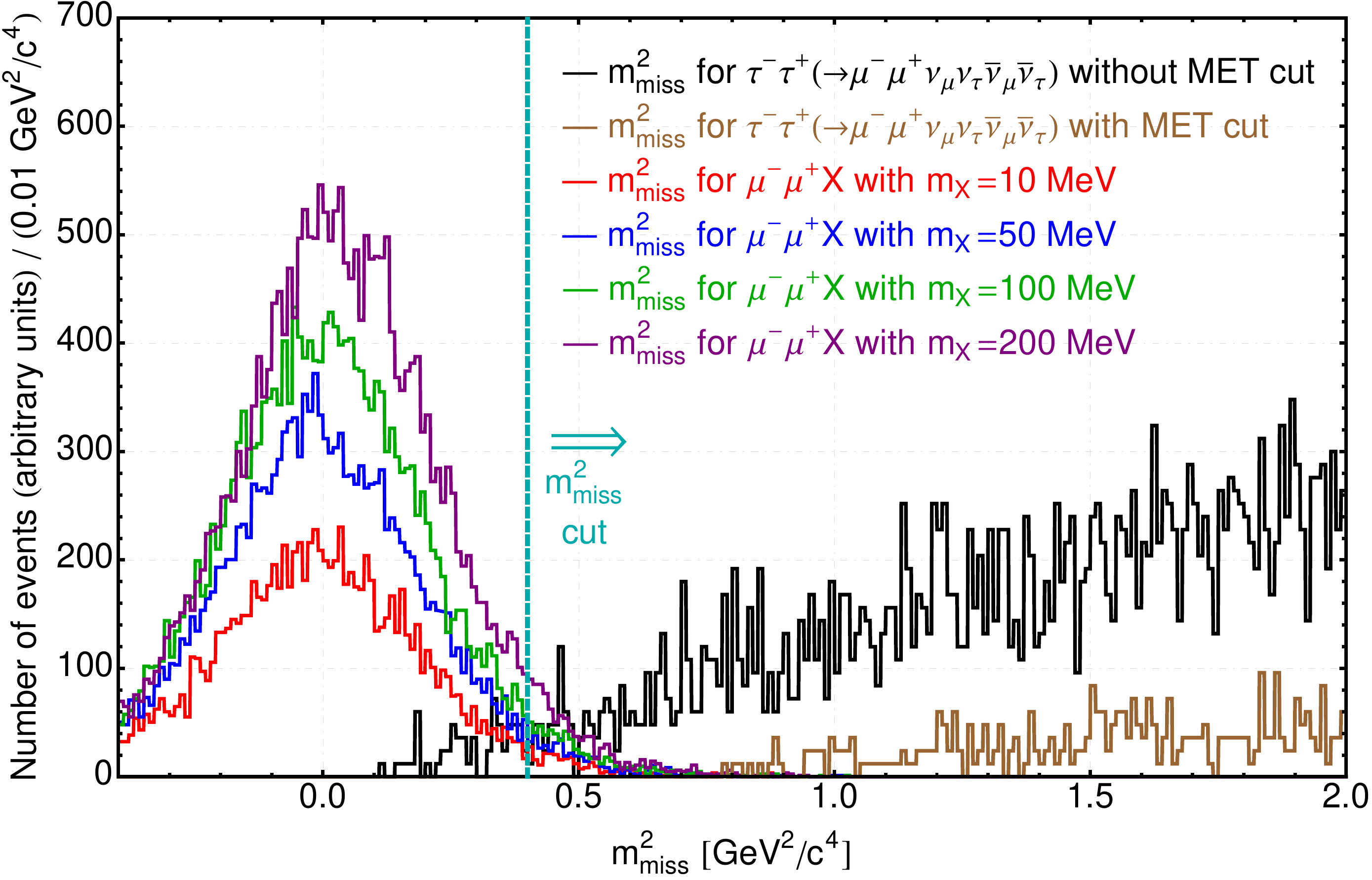}\label{fig_1Dplot_MissingMass_and_MET_cuts}} 
  \caption{$m_{\rm miss}^2$ distribution for muonically decaying $\tau^- \tau^+$ background (with and without missing transverse energy cut $\slashed{E}_T > 1.67 $ GeV) and $\mu^- \mu^+ X$ signal events. All signal events are obtained after $\slashed{E}_T$ cut. We only choose events with $m_{\rm miss}^2 < 0.4$ GeV$^2/c^4$. The missing-mass-squared cut is shown as vertical dotted line. \label{fig_1Dplot_MissingMass_and_MET_cuts}}
\end{figure}
%%%%%%%%%%%%%%%%%%%%%%%%

%%%%%%%%%%%%%%%%%%%%%%%%
%%%%%%%%%%%%%%%%%%%%%%%%

%%%%%%%%%%%%%%%%%%%%%%%%
%%%%%%%%%%%%%%%%%%%%%%%%

\subsection{Sensitivity limit} 

After imposing the kinematic cuts
\begin{eqnarray}
& & i) \ \slashed{E}_T > 1.67 \text{ GeV}, \\ 
& & ii) \ m_{\rm miss}^2 < 0.4 \text{ GeV}^2/c^4
\end{eqnarray}
almost SM backgrounds are removed and the remaining signal $e^- e^+ \rightarrow \mu^- \mu^+ X (\rightarrow \nu \bar{\nu})$ gives the $3\sigma$ sensitivity limit from the criterion
\begin{eqnarray} 
\frac{\mathcal{S}}{\sqrt{\mathcal{S} + \mathcal{B}}} & \geq & 3
\end{eqnarray}
where the signal and the background rates $\mathcal{S}$, $\mathcal{B}$ are given by
\begin{eqnarray}
\mathcal{S} & = & \left [ \int_{\slashed{E}_T^{\rm cut}}^{\sqrt{s}/2} d \slashed{E}_T \int_{(m_{\rm miss}^2)_{\rm cut}^{\min}}^{(m_{\rm miss}^2)_{\rm cut}^{\max}} dm_{\rm miss}^2 \ [ \epsilon (p_{\mu^\pm}) ]^2 \ \frac{d^2 \sigma_{\mu \mu X}^{\rm signal} (s)}{d \slashed{E}_T d m_{\rm miss}^2} \right ] \cdot \int \mathcal{L} \ dt, \\
\mathcal{B} & = & \left [ \int_{\slashed{E}_T^{\rm cut}}^{\sqrt{s}/2} d \slashed{E}_T \int_{(m_{\rm miss}^2)_{\rm cut}^{\min}}^{(m_{\rm miss}^2)_{\rm cut}^{\max}} dm_{\rm miss}^2 \ [ \epsilon (p_{\mu^\pm}) ]^2 \ \frac{d^2 \sigma_{\mu \mu \gamma, \tau \tau, W^*/Z^*}^{\rm background} (s)}{d \slashed{E}_T d m_{\rm miss}^2} \right ] \cdot \int \mathcal{L} \ dt, 
\end{eqnarray}
respectively and $\slashed{E}_T^{\rm cut} = 1.67$ GeV, $(m_{\rm miss}^2)_{\rm cut} = 0.4 $ GeV${}^2 / c^4$ as we mentioned. We also reject all events including muons with momentum below 0.6 GeV$/c$ in the lab frame and assume that the detection efficiency at the $K_L$ and muon (KLM) detector is $\epsilon (p_{\mu^\pm}) = 0.9$ for $p_\pm > 0.6 \text{ GeV}/c$ \cite{Kou:2018nap}. We focus on cases of integrated luminosity $\int \mathcal{L} \ dt= 1, 10, 50$ ab${}^{-1}$. Expected $3\sigma$ sensitivity limits at Belle~II are shown in Fig. \ref{fig_Belle2_mu_philic_sensitivity}. We assume the photon detention inefficiency $1-\epsilon_\gamma = 10^{-6}$ and show other detection inefficiency cases.

\begin{figure}[h]
\centering
\subfloat[$\tau^- \tau^+(\rightarrow \mu^- \mu^+ \nu \nu \bar{\nu} \bar{\nu})$]
{\includegraphics[width=0.318\textwidth]{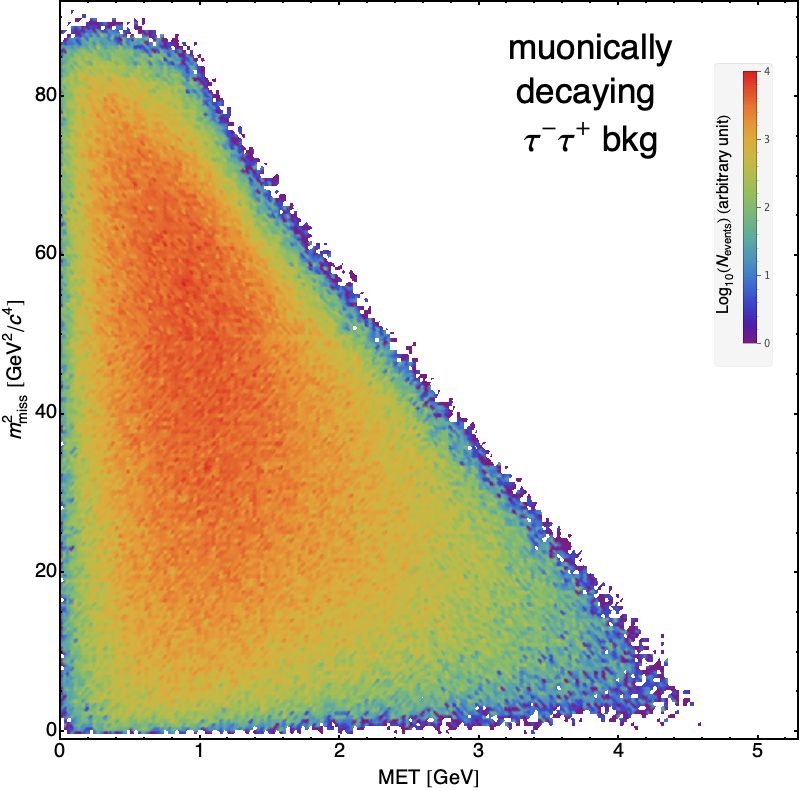}\label{fig_2Dplot_METMM2_bkg03b_muonically_decaying_tautau_HeavyX}} \, 
\subfloat[$\mu^- \mu^+ X$, $m_X = 800$ MeV]
{\includegraphics[width=0.32\textwidth]{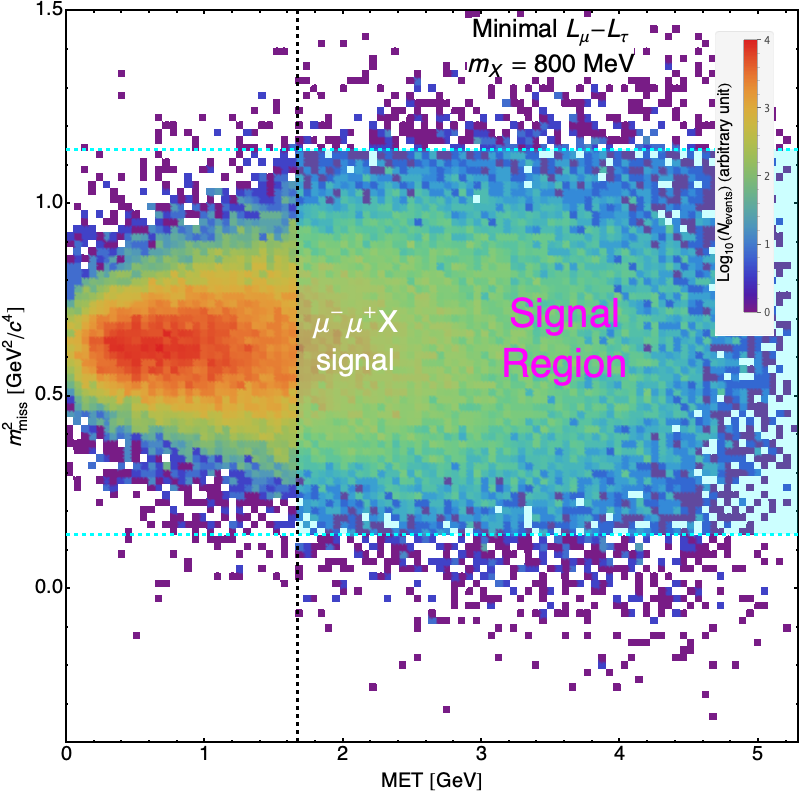}\label{fig_2Dplot_METMM2_sig02_01_minimal_LmuLtau_HeavyX_mX_800MeV}} \, 
\subfloat[$\mu^- \mu^+ X$, $m_X = 3$ GeV]
{\includegraphics[width=0.324\textwidth]{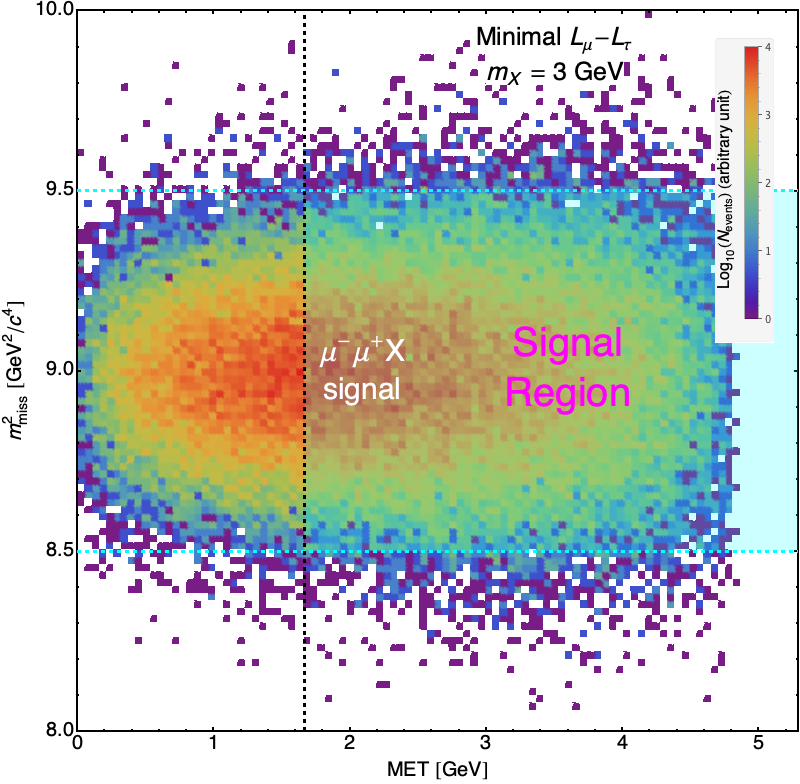}\label{fig_2Dplot_METMM2_sig02_02_minimal_LmuLtau_HeavyX_mX_3GeV}} 
\caption{ Distributions of SM background events $\tau^- \tau^+(\rightarrow \mu^- \mu^+ \nu \nu \bar{\nu} \bar{\nu})$ and $\mu^- \mu^+ X, X \rightarrow (\text{invisible})$ signal events for larger $X$ boson masses ($m_X = 800 \text{ MeV}, \ 3 \text{ GeV}$) on the plane of ($\slashed{E}_T$, $m_{\rm miss}^2$) space.} 
\label{METandMM2_SM_bkg_HeavyX}
\end{figure}

 For $m_X > 2 m_\mu$,  the branching fraction for invisible decays becomes less than the unity, hence reducing the signal rate. For larger values of $X$ boson mass ($m_X \gsim 1$ GeV$/c^2$), the most important background is muonically decaying tau pair ($\tau^+ \tau^- \rightarrow \mu^- \mu^+ \nu_\mu \nu_\tau \bar{\nu}_\mu \bar{\nu}_\tau$) which have large $\slashed{E}_T$ and $m_{\rm miss}^2$. We show the distributions of background and signal events for larger masses of $X$ boson in Fig. \ref{METandMM2_SM_bkg_HeavyX}. We use the kinematic cuts
\begin{eqnarray}
& & i) \ \slashed{E}_T > 1.67 \text{ GeV}, \\ 
& & ii) \ | m_{\rm miss}^2 - m_X^2 | < 0.5 \text{ GeV}^2/c^4
\end{eqnarray}
to obtain the sensitivity limit of the channel $\mu^- \mu^+ + \text{INV}$ for $m_X = 0.5 \sim 8.0 \text{ GeV}^2/c^4$. The sensitivity limit including this larger mass region is shown in Fig. \ref{fig_Belle2_mu_philic_sensitivity_HeavyX}. The sensitivity limit for larger $X$ boson masses do not depend on the photon detection inefficiency, because $\mu^+ \mu^- + \gamma_{\rm unobserved}$ is no longer dominant background for $m_X \gsim 1$ GeV$/c^2$. In this mass region, the best channel is 4-muon mode ($e^- e^+ \rightarrow \mu^- \mu^+ X,\ X \rightarrow \mu^- \mu^+$ as in Ref. \cite{TheBABAR:2016rlg}) due to the huge $\tau^- \tau^+ \rightarrow 2\mu 4\nu$ background for invisibly decaying $X$ case.

Belle II together with NA62 and DUNE are the currently operating or recently approved experiments  that will probe the entire $(g-2)_\mu$ parameter region from $L_\mu-L_\tau$ model with light $X$ boson in the near future. Compared to the kaon decays at NA62 and neutrino-trident process at DUNE, both of which include hadronic amplitude uncertainties,  exploiting the $\mu^+\mu^-X$ signal at Belle II has the merit of less theoretical uncertainties being involved.

%%%%%%%%%%%%%%%%%%%%%%%%
\begin{figure}[h]
\centering
 {\includegraphics[width=0.9\textwidth]{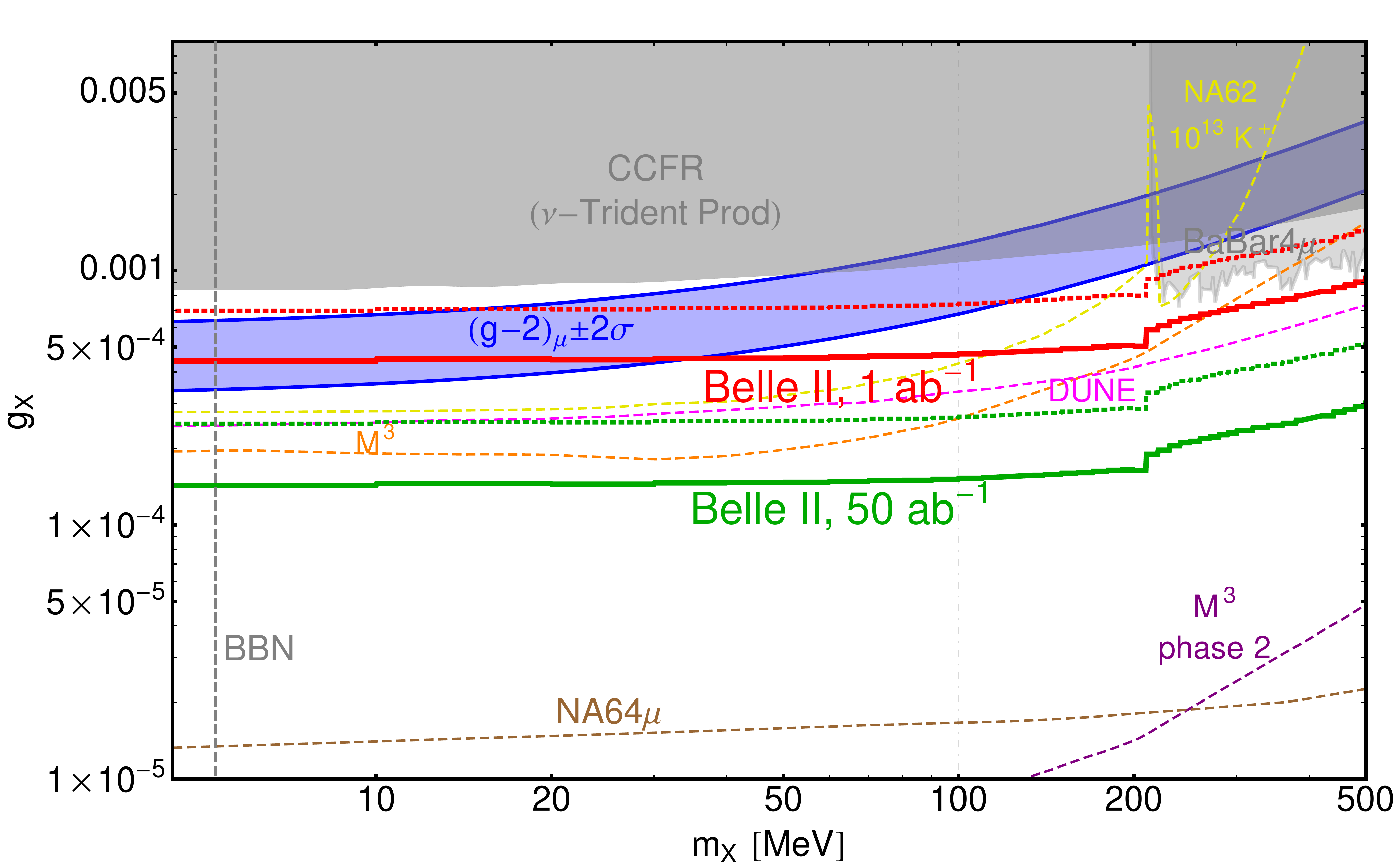}\label{fig_Belle2_mu_philic_sensitivity}} \\
  \caption{Sensitivity limit corresponding to $\mathcal{S}/\sqrt{\mathcal{S}+\mathcal{B}} = 3$ with kinematically optimized signals at $\int \mathcal{L} \ dt = 1, 50\text{ ab}{}^{-1}$ search in the Belle~II experiment with the photon detection inefficiency $1-\epsilon_\gamma = 10^{-6}$ for $\mu^-\mu^+ (\gamma)$ backgrounds (red, green solid lines). We show the case with $1-\epsilon_\gamma = 10^{-5}$ (red, green dotted lines) as the conservative choice of detection inefficiency.  Expected sensitivities of future coming beam-dump experiments such as M$^3$ \cite{Kahn:2018cqs} (orange, purple dashed lines), NA64$\mu$ \cite{Gninenko:2014pea} (brown dashed line), NA62 with $10^{13}$ kaons \cite{Krnjaic:2019rsv} (yellow dashed line) and neutrino trident production in DUNE \cite{Ballett:2019xoj} (magenta dashed line)  are also shown for comparison. \label{fig_Belle2_mu_philic_sensitivity}}
\end{figure}
%%%%%%%%%%%%%%%%%%%%%%%%

%%%%%%%%%%%%%%%%%%%%%%%%
%%%%%%%%%%%%%%%%%%%%%%%%

%%%%%%%%%%%%%%%%%%%%%%%%
\begin{figure}[h]
\centering
 {\includegraphics[width=0.9\textwidth]{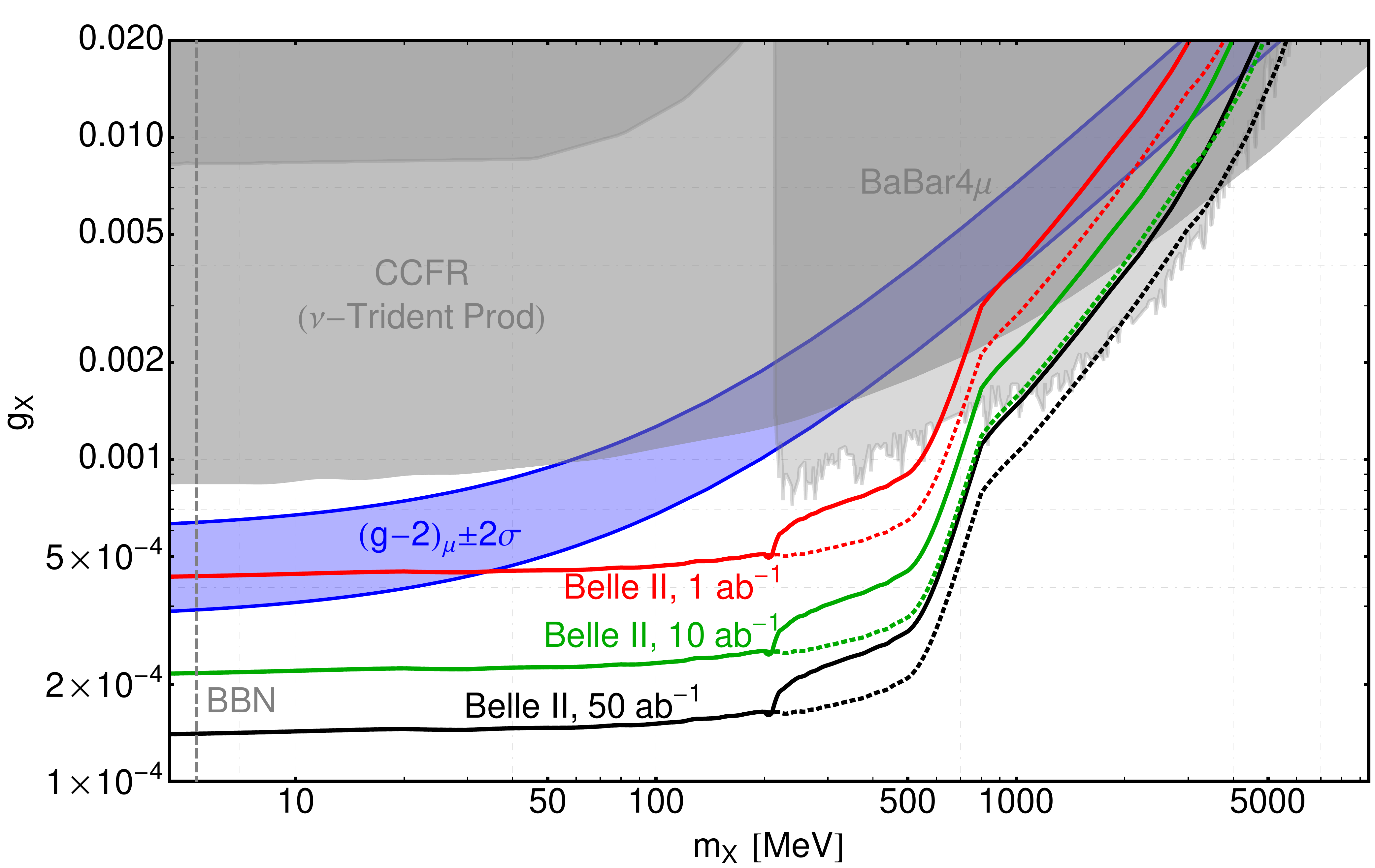}\label{fig_Belle2_mu_philic_sensitivity_HeavyX}} \\
  \caption{ Sensitivity limit, using $\mu^+ \mu^- + \text{INV}$ channel, corresponding to $\mathcal{S}/\sqrt{\mathcal{S}+\mathcal{B}} = 3$ including heavy masses of $X$ gauge boson ($m_X > 2m_\mu$). Red, green, black solid lines are 1, 10, 50 ab${}^{-1}$ search for minimal $L_\mu - L_\tau$ case (which has no additional decay channel to dark sector) and dotted lines indicate Br$(X \rightarrow \text{invisible}) \simeq 1$ cases. \label{fig_Belle2_mu_philic_sensitivity_HeavyX}}
\end{figure}
%%%%%%%%%%%%%%%%%%%%%%%%

\section{Conclusion \label{sec:conclusion}}

The large amount of integrated luminosity is expected in the Belle~II experiment. We expect that the (invisibly decaying) muon-philic light ($m_X \lsim 2 m_\mu$) gauge boson can be probed down to $g_X \gsim 1.5 \times 10^{-4} \ (4.6 \times 10^{-4}, \ 2.3 \times 10^{-4})$ for 50 (1, 10) ab${}^{-1}$ search, rejecting almost SM background events ($\mu^- \mu^+ \gamma$, $\tau^- \tau^+$, $W^*/Z^*$ involved) by imposing $\slashed{E}_T^{\rm cut}$ and $(m_{\rm miss}^2)_{\rm cut}$ simultaneously. This sensitivity limit is largely model-independent. This direct search of muon-philic gauge boson $e^- e^+ \rightarrow \mu^- \mu^+ X$ also can be combined with other channel search, for instance $e^- e^+ \rightarrow \gamma X$, to determine the kinetic mixing $\epsilon_{\gamma X}$ and the fate of $(g-2)_\mu$ explanation scenario by muon-philic light $X$ gauge boson.

%\vspace{1.0cm}
%====================================

\acknowledgments
This work was supported by the National Research Foundation of Korea (NRF) grant funded
by the Korean government (MSIP) (NRF- 2018R1A4A1025334).  PYT was supported by World Premier International Research Center Initiative (WPI), MEXT, Japan.

\bibliographystyle{JHEP}
\bibliography{biblio}

\end{document}